\documentclass[twocolumn,tighten,times,resetfootnote]{aastex7}

\usepackage{xspace}
\usepackage{CJK}
\newcommand{\gkai}[1]{\begin{CJK}{UTF8}{gkai}{#1}\end{CJK}}

\newcommand{\obj}{ZTF\,J0328$-$1219\xspace}

\shorttitle{Variable Debris Transits in \obj}
\shortauthors{Guidry et al.}
\received{August 21, 2026}
\revised{October 26, 2025}
\accepted{November 9, 2026}

\submitjournal{ApJ}

\graphicspath{{./}{figures/}}

\begin{document}
\begin{CJK}{UTF8}{gbsn}

\title{Near-Ultraviolet and Optical Observations of the White Dwarf ZTF\,J0328$-$1219 Reveal Variable Periodicity in its Transiting Debris System}

\author[orcid=0000-0001-9632-7347]{Joseph A. Guidry}
\affiliation{Department of Astronomy \& Institute for Astrophysical Research Boston University Boston, MA 02215, USA}
\email[show]{jaguidry@bu.edu}  

\author[orcid=0000-0001-5941-2286]{J.~J. Hermes}
\affiliation{Department of Astronomy \& Institute for Astrophysical Research Boston University Boston, MA 02215, USA}
\email{jjhermes@bu.edu}

\author[orcid=0000-0002-0853-3464]{Zachary P. Vanderbosch}
\affiliation{Hobby-Eberly Telescope, 32 Mt. Locke Rd., McDonald Observatory, TX 79734, USA}
\affiliation{Department of Astronomy, University of Texas at Austin, Austin, TX 78712, USA}
\email{zachary.vanderbosch@austin.utexas.edu}

\author[0000-0002-2761-3005]{Boris G{\"a}nsicke}
\affiliation{Department of Physics, University of Warwick, Coventry CV4 7AL, UK}
\email{boris.gaensicke@gmail.com}

\author[0000-0002-9632-1436]{Simon Blouin}
\affiliation{Department of Physics and Astronomy, University of Victoria, Victoria, British Columbia V8W 2Y2, Canada}
\email{sblouin@uvic.ca}

\author[orcid=0000-0002-3553-9474]{Laura K. Rogers}
\affiliation{NOIRLab, 950 N Cherry Ave, Tucson, AZ, 85719, USA}
\email{laura.rogers@noirlab.edu} 

\author[0000-0003-1748-602X]{Jay Farihi}
\affiliation{Department of Physics and Astronomy, University College London, London WC1E 6BT, UK}
\email{j.farihi@ucl.ac.uk}

\author[0000-0002-6270-8624]{Stefan Arseneau}
\affiliation{Department of Astronomy \& Institute for Astrophysical Research Boston University Boston, MA 02215, USA}
\email{arseneau@bu.edu}

\author[0009-0002-6065-3292]{Lou~Baya Ould~Rouis}
\affiliation{Department of Astronomy \& Institute for Astrophysical Research Boston University Boston, MA 02215, USA}
\email{lbor@bu.edu} 

\author[orcid=0000-0002-8808-4282]{Siyi Xu (许\CJKfamily{bsmi}偲\CJKfamily{gbsn}艺)} 
\affiliation{NOIRLab, 950 N Cherry Ave, Tucson, AZ, 85719, USA}
\email{siyi.xu@noirlab.edu}

\author[0000-0002-4080-1342]{Bruce Gary}
\affiliation{Hereford Arizona Observatory, Hereford, AZ, USA 85615}
\email{bruce@brucegary.net}

\author[]{Thomas G. Kaye}
\affiliation{Raemor Vista Observatory, Sierra Vista, AZ, USA 85650}
\email{tom@TomKaye.com}

\author[0009-0004-7915-2775]{Yarin-Meir Shani}
\affiliation{Department of Particle Physics and Astrophysics, Weizmann Institute of Science, Rehovot 7610001, Israel}
\email{yarin-meir.shani@weizmann.ac.il}

\author[0000-0003-0496-8732]{Hannah Worters}
\affiliation{South African Astronomical Observatory, P.O Box 9, Observatory 7935, Cape Town, South Africa}
\email{h.worters@saao.nrf.ac.za}

\collaboration{24}{Whole Earth Telescope Observers}

\author[0009-0001-6853-9470]{Kaylee E. Grace}
\affiliation{Department of Physics \& Astronomy, 217 Sharp Lab, University of Delaware, Newark, DE 19716, USA}
\email{kgrace@udel.edu}

\author[0009-0003-5777-092X]{Steven Z. Savery}
\affiliation{Department of Physics \& Astronomy, 217 Sharp Lab, University of Delaware, Newark, DE 19716, USA}
\email{savery@udel.edu}

\author[0000-0002-7792-0492]{J. L. Provencal}
\affiliation{Department of Physics \& Astronomy, 217 Sharp Lab, University of Delaware, Newark, DE 19716, USA}
\affiliation{Mount Cuba Astronomical Observatory, Greenville, DE 19807, USA}
\email{jlp@udel.edu}

\author[0000-0002-0933-6403]{Chun-Hai Bai (\gkai{白春海})}
\affiliation{Xinjiang Astronomical Observatory, Chinese Academy of Sciences, Urumqi, 830011, People's Republic of China}
\email{baichunhai@xao.ac.cn}

\author[0000-0002-0656-032X]{Keaton J. Bell}
\affiliation{Department of Physics, Queens College, City University of New York, Flushing, NY-11367, USA}
\email{Keaton.Bell@qc.cuny.edu}

\author[0000-0001-7010-7637]{Barbara G. Castanheira}
\affiliation{Department of Physics and Astronomy, Baylor University, Waco, TX 76798-7316, USA}
\email{Barbara_Endl@baylor.edu}

\author[0009-0006-2789-0893]{Paul Chote}
\affiliation{Department of Physics, University of Warwick, Coventry CV4 7AL, UK}
\affiliation{Centre for Space Domain Awareness, University of Warwick, Gibbet Hill Road, Coventry CV4 7AL, UK}
\email{p.chote@warwick.ac.uk}

\author[]{Millie A. Dill}
\affiliation{Department of Physics \& Astronomy, 217 Sharp Lab, University of Delaware, Newark, DE 19716, USA}
\email{mdill@udel.edu}

\author[]{Inajara Endl}
\affiliation{Cedar Park High School, Cedar Park, TX 78613, USA}
\email{ina.cendl@gmail.com}

\author[]{Torben Endl}
\affiliation{Cedar Park High School, Cedar Park, TX 78613, USA}
\email{Torben.endl@gmail.com}

\author[0000-0002-2781-873X]{Lester Fox~Machado}
\affiliation{Universidad Nacional Aut{\'o}noma de M{\'e}xico, Instituto de Astronom{\'i}a,  A.P. 106, 22800, Ensenada, B.C., M{\'e}xico}
\email{lfox@astro.unam.mx}

\author[0000-0003-3688-2298]{Elizabeth M. Green}
\affiliation{Steward Observatory, University of Arizona, 933 N. Cherry Avenue, Tucson, AZ 85721, USA}
\email{egreen@arizona.edu}

\author[]{Kim Green}
\affiliation{Mount Cuba Astronomical Observatory, Greenville, DE 19807, USA}
\email{kimgreenmcao@gmail.com}

\author[0000-0003-1202-3683]{Easton J. Honaker}
\affiliation{Department of Physics \& Astronomy, 217 Sharp Lab, University of Delaware, Newark, DE 19716, USA}
\email{ehonaker@udel.edu}

\author[0009-0007-9276-9886]{Nikoo Hosseininezhad}
\affiliation{City University of New York Graduate Center, 365 5th Ave, New York, NY 10016, USA}
\email{shosseininezhad@gradcenter.cuny.edu}

\author[]{Xiao-Jun Jiang (\gkai{姜晓军})}
\affiliation{National Astronomical Observatories, Chinese Academy of Sciences, Beijing 100101, People's Republic of China}
\affiliation{University of Chinese Academy of Sciences, Beijing 100049, People's Republic of China}
\email{xjjiang@nao.cas.cn}

\author[0000-0002-7562-1675]{Momin Y. Khan}
\affiliation{Department of Physics and Astronomy, Baylor University, Waco, TX 76798-7316, USA}
\email{Momin_Khan1@baylor.edu}

\author[0009-0004-6130-7775]{Morgan A. Mitchell}
\affiliation{Department of Physics, University of Warwick, Coventry CV4 7AL, UK}
\affiliation{Centre for Exoplanets and Habitability, University of Warwick, Gibbet Hill Road, Coventry CV4 7AL, UK}
\email{morgan.mitchell@warwick.ac.uk}

\author[0000-0002-6748-1748]{Michael H. Montgomery}
\affiliation{Department of Astronomy, University of Texas at Austin, Austin, TX 78712, USA}
\email{mikemon@astro.as.utexas.edu}

\author[0009-0006-1574-5197]{Rudy Morales}
\affiliation{Department of Physics and Astronomy, Baylor University, Waco, TX 76798-7316, USA}
\email{Rudy_Morales1@baylor.edu}

\author[0000-0002-7703-6701]{Tali Natan}
\affiliation{Department of Physics \& Astronomy, 217 Sharp Lab, University of Delaware, Newark, DE 19716, USA}
\email{tali@udel.edu}

\author[0000-0003-3271-9709]{Hui-Juan Wang (\gkai{王汇娟})}
\affiliation{National Astronomical Observatories, Chinese Academy of Sciences, Beijing 100101, People's Republic of China}
\affiliation{University of Chinese Academy of Sciences, Beijing 100049, People's Republic of China}
\email{wanghj@nao.cas.cn}

\author[]{Jia-Qi Wang (\gkai{王佳琪})}
\affiliation{National Astronomical Observatories, Chinese Academy of Sciences, Beijing 100101, People's Republic of China}
\email{wangjiaqi@nao.cas.cn}

\author[]{Tu-Hong Zhong (\gkai{钟土红})}
\affiliation{Xinjiang Astronomical Observatory, Chinese Academy of Sciences, Urumqi, 830011, People's Republic of China}
\email{zhongtuhong@xao.ac.cn}

\begin{abstract}

We observed the transiting debris disk system around the white dwarf \obj using the Hubble Space Telescope (HST) in 2025~October, obtaining near-ultraviolet (NUV) spectroscopy with STIS over five consecutive HST orbits. Our abundance analysis supports the accretion of a chondritic body. We complemented our HST program with 16 days of optical time-series photometry beginning in 2025~October, which reveal variable periodicity and rapidly evolving transit morphology. Previous monitoring by TESS during Sectors~4 and 31 confirmed a predominant, stable transit orbital period, $P_{\rm A} = 9.937$\,hr, along with a second, transient period, $P_{\rm B} = 11.2\,$hr. Our 2025~autumn light curves show transits are dominated by $P_{\rm B} = 11.17$\,hr with $P_{\rm A}$ initially appearing as a weaker, secondary feature and eventually returning to predominance. We also identify a candidate transit period of 7.64\,hr during the final days of the 2025 autumn campaign, although its significance and persistence remain uncertain. We add simultaneous $ugr$ light curves collected with ULTRACAM in 2024~November that are dominated by $P_{\rm A}$ and show no significant color term. We further compare the transit depths in optical photometry concurrent with our NUV observations, again measuring a lack of transit-dependent extinction except during the fifth HST orbit where a reddening term may arise. The observed variable periodicity and potentially changing color behavior illuminate ongoing dynamical processes, namely collisions, and demonstrate the opportunity for studying dust physics in real time at white dwarf debris disks.

\end{abstract}

\keywords{\uat{White dwarf stars}{1799} --- \uat{Transits}{1711} --- \uat{Period determination}{1211} --- \uat{Debris disks}{363} --- \uat{Circumstellar dust}{236} --- \uat{Exoplanet systems}{484}}

\section{Introduction}\label{sec:intro}

Both bookends of stellar evolution are influenced by circumstellar debris. Protostars are born embedded in thick, flared disks that nurse planetary formation (see \citealt{White2007,Williams2011}, and the references therein). Spectroscopy of white dwarf stars indicates some of this material that coagulated into planets and planetesimals is recycled back into circumstellar debris disks many eons later. More than 40\% of white dwarf spectra show photospheric absorption of heavy metals, indicating the recent accretion of planetary material \citep{Zuckerman2003,Zuckerman2010,Koester2014,OuldRouis2024}.

The debris white dwarfs accrete arrives at the circumstellar environment following the dynamical scattering of a minor planetary body by at least one larger, intact planet that survived post-main sequence evolution. Upon entering the tidal disruption radius of the star on a likely extremely eccentric orbit, the material is disrupted and ground down into an eccentric debris disk that circularizes and migrates inwards towards the star \citep{Debes2002,Jura2003}. Eventually, sufficiently fine dust is sublimated and then dragged onto the stellar photosphere, manifesting the observed metal pollution (see reviews by \citealt{Farihi2016,Veras2024}, and the references therein).

White dwarfs are observed to accrete from their remnant planetary systems across the single-star evolution cooling sequence \citep{Blouin2022}. Accretion is most sensitively detected in the ultraviolet, where recent accretion can be assessed through the relative abundance ratios of silicon and carbon \citep{Koester2014,OuldRouis2024}. X-ray emission from G\,29$-$38 directly verifies the bombardment accretion of planetary debris \citep{Cunningham2022}.

Derived relative abundances from UV and optical spectra are generally consistent with bulk Earth or chondritic natures \citep{Gaensicke2012,Xu2014,Xu2019b,XuBonsor2021,Bonsor2026}, and on rare occasion are consistent with water-rich parent bodies \citep{Farihi2013,Trierweiler2025}. UV spectroscopy is often required to obtain the most-accurate census of pollutants, even for cooler white dwarfs \citep{Hollands2022}, though systematic discrepancies between the derived abundances from the optical and the ultraviolet persist \citep{Gaensicke2012,Xu2019b,LeBourdais2024}. Mid-infrared spectroscopy illuminates the mineralogy of the debris, where 10-$\mu$m silicate emission from the disk dominates \citep{Farihi2025,Rogers2025b}.

White dwarf debris disks are ubiquitously dynamic \citep{Noor2025}, having their emitting areas and particle size distributions regularly reshaped by collisional cascades \citep{KenyonBromley2017a,KenyonBromley2017b,Swan2021}. Like their protoplanetary counterparts, debris disks around white dwarfs are likely flared and fall along random inclinations \citep{Jura2007,Ballering2022,Bhattacharjee2025b}. The systems with the most edge-on inclinations should cast dimmings as the debris transits the stellar cross section. More than 20 transiting debris systems are now known \citep{VanRoestel2026,OuldRouis2026b}, with most hosting the debris well beyond the nominal Roche limit \citep{Veras2026}. All but five transiting debris systems appear metal polluted in low-resolution spectroscopy--all of five which likely travel extremely eccentric, long-duration orbits \citep{Guidry2021,VanRoestel2026}--and at least two show infrared excesses: WD\,1145+017 \citep{Xu2018} and SBSS\,1232+563 \citep{Debes2011,Swan2020}.

Transit recurrence timescales indicate a distribution of orbits. There are at least two cases of debris orbiting near the tidal disruption radius of the host star, WD\,1145+017 at 4.49\,hr \citep{Vanderburg2015} and ZTF\,J1944+4557 at 4.9704\,hr \citep{Guidry2025}, next to the 5.910\,hr period that briefly emerged during a deep transit of ZTF\,J0436+5728 \citep{VanRoestel2026}. Three others show orbital periods ranging from about half a day to one day: ZTF\,J0328$-$1219 (9.937\,hr and 11.2\,hr; \citealt{Vanderbosch2021}), SBSS\,1232+563 (14.842\,hr; \citealt{Hermes2025}), and WD\,1054$-$226 (25.02\,hr; \citealt{Farihi2022}). ZTF\,J0139+5245 enters deep transits roughly every 107\,days \citep{Vanderbosch2020}. The probability distribution of orbital periods for tidally disrupted asteroids around white dwarfs matches this spread, roughly covering $4\,{\rm hr} \lesssim P_{\rm orb} \lesssim 10^2$\,d and peaking over 12--16\,hr \citep{Li2025}.

Recurrence timescales are unmeasured at all other transiting systems. Several systems have shown only a single transit in their ZTF light curve across over eight years of monitoring, including ZTF\,J0347$-$1802 \citep{Guidry2021},  ZTF\,J1013$-$0427 \citep{Bhattacharjee2025a}, and five systems published in \citet{VanRoestel2026}. SBSS\,1232+563 enters deep transits sporadically every 6-8 years, in addition to the 14.8\,hr periodicity observed by TESS following the 2023 deep transit \citep{Hermes2025}. ZTF\,J0923+4236 undergoes deep transits every few weeks without clear periodicity \citep{Guidry2021,Aungwerojwit2024}. Several systems that show transits on short timescales have yet to be verified to show periodic transits \citep{Guidry2021,Bhattacharjee2025a}.

ZTF\,J0328$-$1219 (Gaia~DR3 5161807767825277184) was confirmed as a spectral type DZ white dwarf host to transiting debris by \citet{Guidry2021}. Periodic transits repeating at $P_{\rm A} = 9.937\,$hr were observed in follow-up light curves collected during 2020~December and 2021~February and corroborated by photometry from TESS Sectors~4 and 31 \citep{Vanderbosch2021}. A second period, $P_{\rm B} = 11.2$\,hr, was detected exclusively during the first half of TESS Sector~31 in 2020~October. High-resolution spectroscopy revealed circumstellar gas absorbing in the optical at the sodium D doublet and in the near-infrared at the central line of the calcium II triplet ($\lambda = 8542$\,\AA; \citealt{Vanderbosch2021}). Evidence for circumstellar dust around \obj in the near-infrared is lacking. Photometry from unWISE of the system are consistent with the white dwarf model photosphere, although there is potentially significant excess emission at the $J$-band \citep{Vanderbosch2021}. The system is not observed to vary in its unWISE light curve collected by NEOWISE \citep{Guidry2024}\footnote{See the supplemental catalog available at \citet{Guidry2024_WISE_Zenodo}}. Optical survey photometry records evidence for transits occurring since at least 2006, including a dimming of roughly 0.3\,mag beginning in 2012 and lasting through 2014 \citep{Aungwerojwit2024}. The transits were consistent with being colorless across both $g$- and $r$-band ZTF DR5 photometry \citep{Vanderbosch2021} and follow-up $g$- and $R$-band light curves \citep{Gary2024}.

\begin{figure*}[!t]
    \epsscale{1.15}
    \plotone{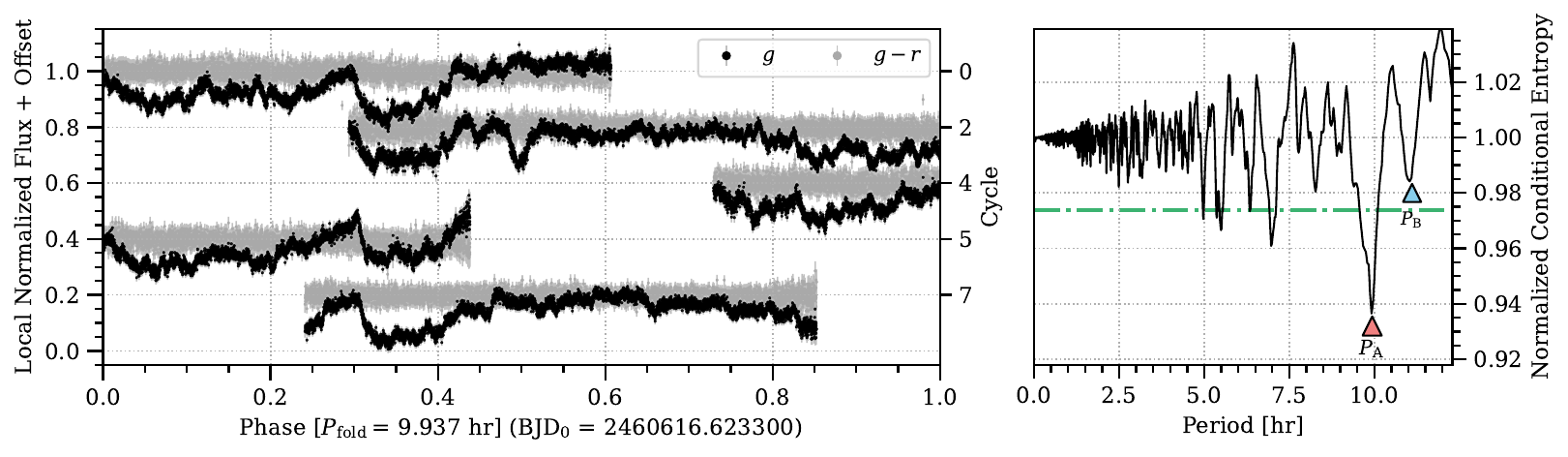}
    \caption{2024~November 2-5 ULTRACAM $g$-band (black) and $g{-}r$ (gray) light curves phase folded on $P_{\rm A}=9.937$\,hr (left) and the conditional entropy periodogram of combined, averaged $g$ \& $r$ light curve (right). A prominent transit centered around Phase$\,\approx\,$0.38 persists across the four consecutive nights. We annotate on the periodogram $P_{\rm A}$ and $P_{\rm B}$ as red and light blue triangles, respectively. Only the A period is significant to the nominal $10\,\rho$ threshold (see Section~\ref{sec:CE}), delineated by the dash-dotted green line. The disappearance of features in the flat $g{-}r$ light curve indicates achromatic transits (Section~\ref{sec:ucam_colors}).
    }\label{fig:UCAM_Nov2024_pfold_LCs}
\end{figure*}

In this manuscript we publish follow-up, time-series photometry and spectroscopy of \obj. These include Hubble Space Telescope (HST) spectroscopy and time-series photometry in the near-ultraviolet (NUV), and also ground-based, optical observations. Our work tightens constraints on the relative abundances of the accreted debris, confirms consistent transit depths across simultaneous $u$-, $g$-, and $r$-band photometry in 2024~November and between near-concurrent NUV and optical photometry in 2025~October, and demonstrates variable periodicity in the system with the predominant orbital period changing from $P_{\rm A}$ in 2024~November to $P_{\rm B}$ in 2025~October. In Section~\ref{sec:observations} we recount our various observing campaigns. We demonstrate our periodicity analysis in Section~\ref{sec:periodanalysis}. Section~\ref{sec:color} provides our tests for transit color terms. We update the model spectrum of the system in Section~\ref{sec:model_spec}. We contextualize these results and offer explanations for the observed dynamical activity in Section~\ref{sec:discussion} and recapitulate our key findings in Section~\ref{sec:conclusion}.

\section{Observations of \obj} \label{sec:observations}

\subsection{Optical, Ground-Based, Time-Series Photometry}\label{sec:photfollowup}

\begin{figure*}[!t]
    \epsscale{1.05}
\plotone{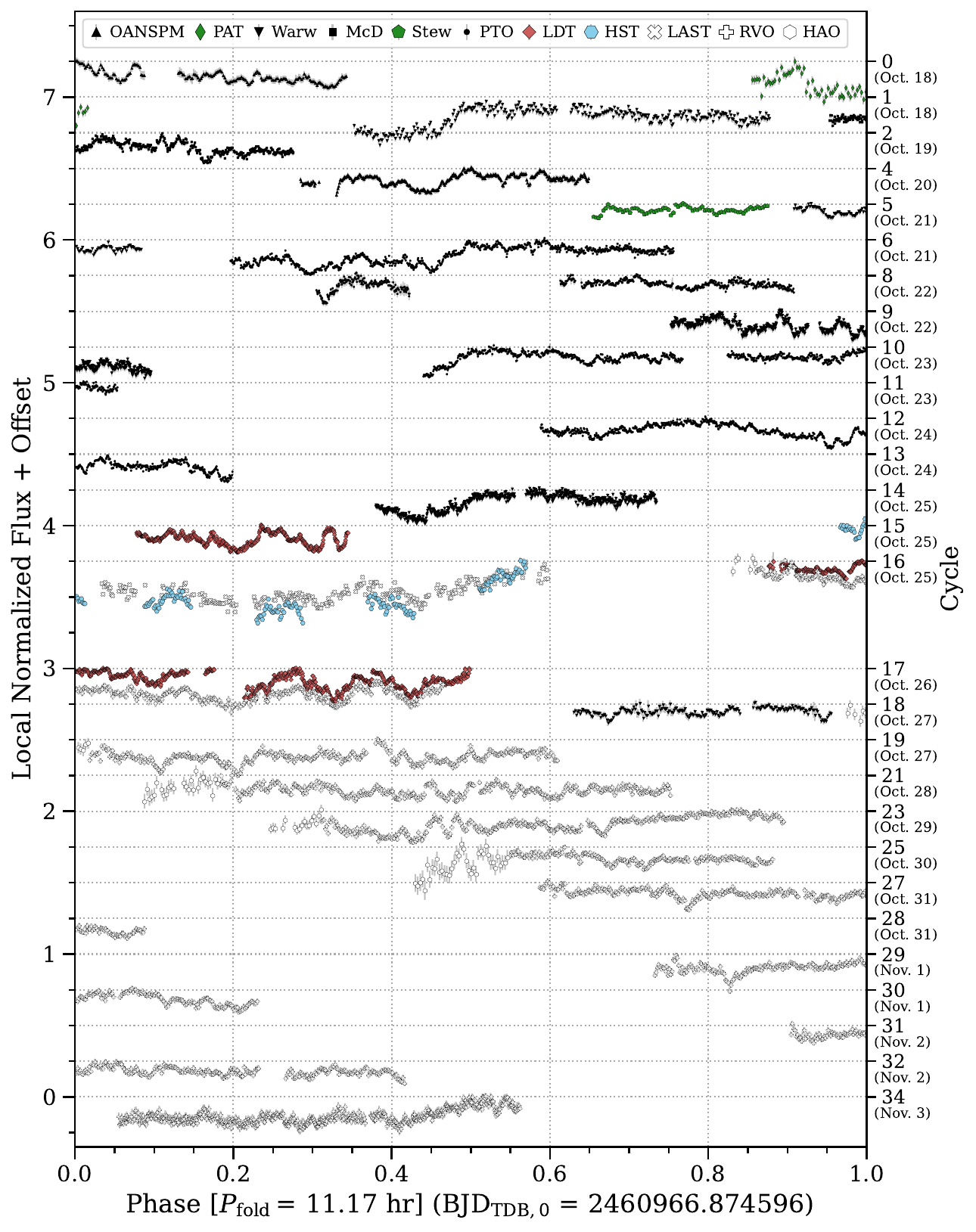}
    \caption{2025~October and November light curves of \obj phase-folded on $P_{\rm B} = 11.17$\,hr, the best measured period for this epoch (Section~\ref{sec:periodanalysis}). Light curves are color-coded by bandpass ({\em BG40}: black, $g$ \& $V$: green, $r$: red, NUV: light blue, and unfiltered: white) and each observatory is assigned a unique marker shape. Repeating patterns are detectable early on, such as at Phase\,$\approx$\,0.5 during Cycles 1, 4, and 6. The transit morphologies evolve much more rapidly at later cycles and no longer show obvious repeating patterns at this period. This dynamical activity is corroborated by the changing periodograms in the bottom panel of Figure~\ref{fig:periodograms}. The offset between Cycles 16 and 17 is boosted to account for the enhanced total flux excursion observed by HST and LAST. Some light curves are binned by a factor of two or three for improved visibility.
    }\label{fig:Oct2025_pfold_LCs}
\end{figure*}

We coordinated several follow-up campaigns for optical, ground-based, high-speed, time-series photometry of \obj, especially to complement our HST program. The methods by which we reduced these photometric data and extracted light curves varied depending on the facility. We describe these campaigns and associated reductions chronologically by the observing date. Our observing log of these photometry is available in Table~\ref{tab:phot_obs}. We plot these light curves in both Figure~\ref{fig:UCAM_Nov2024_pfold_LCs} (2024~November ULTRACAM) and Figure~\ref{fig:Oct2025_pfold_LCs} (2025~autumn light curves).

While reduction methods varied from telescope to telescope, we did not normalize any of our light curves to a low-order (linear or quadratic) polynomial because of our inherent uncertainty of the light curve morphology---we never definitively observed the stable, out-of-transit flux of \obj. Low-order polynomial division corrections are usually performed to correct for differential extinction introduced by the airmass of the target varying over the observing baseline \citep[see, e.g.,][]{Provencal2009}.

{\it ULTRACAM} --- We observed \obj for a total 26.39\,hr across the four consecutive nights of 2024~November~2-5 using the ULTRACAM \citep{Dhillon2007} photometer mounted onto the 3.58-m New Technology Telescope at La~Silla Observatory in Chile. This instrument is a triple-beam, frame-transfer CCD imager. During each night the target and two or more field stars were observed simultaneously in bandpasses similar to the SDSS $ugr$ filters but with higher throughput. Individual frame times were 3.47\,s in all three bands, but in the $u$-band three frames were co-added to improve signal-to-noise (effective exposure times of $10.40$\,s). We show both the $g$-band and residual $g-r$ light curves in Figure~\ref{fig:UCAM_Nov2024_pfold_LCs}. For our analysis of the color term to these transits (Section~\ref{sec:ucam_colors}), we binned the $g$- and $r$-band light curves by a factor of three so we can compare to the $u$-band photometry with consistent exposure times.

Following \citet{Farihi2024}, we processed the raw images using the reduction software {\tt hipercam}\footnote{\url{https://github.com/HiPERCAM/hipercam}} \citep{Dhillon2021}. Each frame was corrected for bias and flat fielded using sky flats taken each night, with typical read noise in the range 2--3\,ADU in each of the three bias frames. We carried out differential photometry on the science target and two field stars for reference.  The reduction package tracks the mean full
width at half maximum for all three stars, and scales photometric apertures to twice this size, with sky annuli that are typically four to eight times larger.  Each light curve was constructed from the science target flux divided by the sum of these two reference stars, normalized to a median of one. We applied barycentric corrections to the timestamps using the methods of \citet{Eastman2010}.

{\it Whole Earth Telescope (WET)} --- We utilized the WET network during the XCOV31 campaign\footnote{\url{https://whitedwarfsarecool.github.io/wet.html}} to observe \obj\ as a secondary target at several sites across the globe in late 2025~October. These included the Ruca imaging system aboard the 1.5-m Harold Johnson Telescope at Observatorio Astron{\'o}mico Nacional at San Pedro M{\'a}rtir (OAN-SPM; \citealt{Zazueta2000}; October~18, 20, 21), the 0.43-m Photometric Auxiliary Telescope at Xingming Observatory using the QHYCCD Cameras2 Capture instrument (PAT; October~18), the Warwick 1m Telescope at Roque de los Muchachos Observatory with the Andor iKon-L CCD camera (Warw; \citealt{Gaensicke2016}; October~18, 21, 23, 25, 27), McDonald Observatory 2.1-m Otto Struve Telescope with the ProEM photometer (McD; October~19), the 1.54-m Kuiper Telescope at the Catalina Station of Steward Observatory with the Mont4K CCD imager (Stew; October~21). Exposure times varied from $10-60$\,s across these sites, aggregating to a total non-overlapping baseline of 35.05\,hr. The OAN-SPM, Warw, and McD observations used a {\em BG40} filter, while the PAT and Stew observations were collected through a $g$- and $V$-band filter, respectively. 

We reduced the WET photometry identically, regardless of facility, following the procedure developed by \citet{Provencal2012}. We used Maestro \citep{Dalessio2010} to perform aperture photometry with circular annuli ranging between 2 and 20 pixels from the centroid of the target, as well as for up to 29 field stars in the images to use as comparisons. We then used {\tt phot2lc} \citep{phot2lc} to create the final divided light curves by removing outliers, optimizing comparison star selections, and applying barycentric corrections to the GPS timestamps using {\tt Astropy} \citep{astropy:2013,astropy:2018,astropy:2022}.

{\it Perkins Telescope Observatory (PTO)} --- At the 1.8-m PTO we observed \obj using the PRISM instrument \citep{Janes2004} over the four consecutive nights of 2025~October~21-24 for a total of 23.90\,hr. All our photometry were collected using the blue-transmitting, red-cutoff {\em BG40} filter with integration times of 20\,sec with windowed chips to 410${\times}$380 pixels at a plate scale of 0.39\arcsec per pixel, reducing the overhead from readouts to typically 5\,s. 

We used custom-built Python-based routines to calibrate our raw science images by subtracting CCD biases and dome flats. We then reduced our images using the {\tt hipercam} pipeline, doing PSF photometry with {\tt variable} apertures and PSF profiles, all conducted in the {\tt optimal} extraction mode \citep{Naylor1998}. To account for sky backgrounds we drew circular annuli 20$-$30\,pixels from the centroids of our target and chosen comparison stars. To optimize our photometry we modified particular parameters, mainly {\tt search\_half\_width}, if the sky conditions reduced the S/N enough to hamper centroid retrieval. We generated divided light curves using the {\tt phot2lc}, clipping 5$\sigma$ outliers and applying barycentric corrections to the timestamps.

{\it Lowell Discovery Telescope (LDT)} --- We immediately followed our PTO campaign with two more consecutive nights of high-speed, time-series photometry on 2025~October~25-26 using the Large Monolithic Imager \citep{Massey2013} 4.3-m LDT. We observed for a total of 9.23\,hr over the two nights with 15\,s exposures in the $r$-band. We windowed the CCD to 500${\times}$500\,pixel arrays with 3${\times}$3 binning yielding a plate scale of 0.36\arcsec\ per pixel. Readout times averaged to roughly 4.55\,s. Our observations on October~26 were heavily impacted by intermittent clouds.

We reduced our LDT/LMI photometry very similarly to the PTO/PRISM photometry, using {\tt hipercam} in the same manner with the sole change of using the {\tt normal} extraction mode instead of {\tt optimal}. We used {\tt phot2lc} to generate optimized light curves with barycentric-corrected timestamps.

{\it Large Array Survey Telescope (LAST)} --- We used three 28-cm telescopes within the LAST array \citep{Ofek2023} to observe \obj on 2025~October~25, providing us with 6.30\,hr of unfiltered optical photometry concurrent with our HST program (Section~\ref{sec:hubble_obs}).

We reduced the images and performed PSF photometry using the standard LAST pipeline \citep{Ofek2023b} one telescope at a time and then coadded the resultant photometry. We carried out absolute photometric calibration following the method of \citet{Garrappa2025}, which models the instrumental response and atmospheric transmission for each image. We applied an additional relative photometric calibration of our target using the comparison star Gaia~DR3 5161911602954639232, which has no Gaia variability flags nor is observed to vary in ZTF, as a comparison star to calibrate our PSF photometry. We applied barycentric corrections to the LAST timestamps using {\tt Astropy}.

{\it Hereford Arizona Observatory (HAO)} --- We observed \obj with the HAO 11-in Celestron telescope based in Hereford, Arizona, the nights of October~26, 27, 28, 30, and 31, exposing for 111\,s without a filter. Readouts were typically about 10\,s. We used MaxIm~DL to calibrate our science images and a custom Excel created by author B.L.G. to optimize the MaxIm~DL photometry. We used {\tt Astropy} to apply barycentric corrections.

{\it Raemor Vista Observatory (RVO)} --- We observed \obj with the RVO 1.1-m telescope based in Sierra Vista, Arizona, the nights of October~26-31 and November~1-3, exposing for 60\,s on all nights except November~3 where we took 20\,s integrations, with typical overheads of roughly 7\,s. Every light curve was collected without a filter. We used AstroImageJ to calibrate and analyze our science images and applied barycentric corrections to the timestamps using {\tt Astropy}.

\subsubsection{Light Curve Normalization}\label{sec:lc_norm}

The true, out-of-transit flux baseline of \obj is unconstrained because we continuously observed it to be transited and we do not know the optical depth of the transiting debris. The {\tt phot2lc} software outputs mean-normalized light curves in units of relative flux, which we notate here as $f_0$. We work in this relative flux space for our period search analysis (Section~\ref{sec:periodanalysis}). 

For our color analysis and light curve presentation, we sought to define a zero point for each individual light curve. We re-normalize the relative fluxes to a factor we define $\bar{f}$:
\begin{equation}
    f_{\rm norm} = \frac{f_0}{\bar{f}}
\end{equation}
Here, $\bar{f}$ is equal to the median of the relative flux values greater than the percentile corresponding to $100-p$, where $p$ is the average point-to-point scatter of the light curve in percent. For example, if a light curve has an average point-to-point scatter of 5\%, then $\bar{f}$ is the median of all fluxes brighter than the 95th percentile of the light curve. The typical point-to-point scatter levels of our light curves range from about $1-5$\% (see the observing log in Table~\ref{tab:phot_obs}). We plot light curves in $f_{\rm norm}$ values as ``Local Normalized Flux.''

\subsection{Hubble Space Telescope/STIS NUV Spectroscopy}\label{sec:hubble_obs}

We targeted \obj with HST in Cycle 32 program GO~\#17789 \citep{2024hst..prop17789G}, obtaining five near-ultraviolet (NUV) spectra over five consecutive HST orbits beginning 2025~October~25 18:54:27 and lasting through 2025~October~26 02:09:02 (all times UTC). We designed our experiment to utilize the NUV MAMA detector in the Space Telescope Imaging Spectrograph (STIS) with the G230L grating and the 52$\times$0.5\arcsec\ aperture in {\sc time-tag} mode centered on $\lambda = 2376\,$\AA\ with an average dispersion of $\Delta \lambda = 1.58$\,\AA/pixel ($R \approx 1500$). Pointing was nominal for each orbit, with RMS jitter $<$0.003\arcsec\ and no pointing excursions $>$0.006\arcsec\ from the target, allowing us to confidently infer slit losses were minimized. The respective integration times, discounting overheads, for each orbit were 1940, 2505, 2482, 2482, and 2482\,s.

For our spectroscopic analysis (Section~\ref{sec:model_spec}), we utilize the Hubble Advanced Spectral Product (HASP; \citealt{2024cos..rept....1D}) co-added and absolute flux-calibrated spectrum available online at MAST.

\subsubsection{NUV Light Curve Extraction}

We generated an NUV light curve from our time-tagged STIS spectroscopy by performing aperture photometry on 2D spectra images generated with 60-s effective integrations. Relying on the guidelines provided by the STIS Data reduction guide \citep{2024stii.book....8R} and STIS Time-Tag Analysis guide \citep{2000stis.rept....2D}, we first partitioned the spectra into 60-s effective exposures using {\tt lightcurve} \citep{Ely2015,Sandhaus2016}. We applied barycentric corrections to the timestamps using the {\sc IRAF}-based routine {\tt ODELAYTIME} as implemented in {\sc PyRAF} \citep{PyRAF2012}. We generated {\em accum} images with 60-s exposures using {\tt calstis} as implemented by {\tt stistools}\footnote{\url{https://stistools.readthedocs.io/en/latest/index.html}}.

We performed aperture photometry on the background-subtracted, flat-fielded {\em flt} images outputted by {\tt calstis} using {\tt photutils} \citep{bradley_2026_19636730}. We drew rectangular apertures to sample the spectrum from 1800-3100\,\AA, yielding dimensions of 838\,pixels wide by 10\,pixels in height. The target spectrum in the calibrated 2D spectra tends to be Gaussian with a typical standard deviation of $1.5 - 2.0$\,pixels, hence our choice for a height of 10\,pixels to sample roughly $\pm$5$\sigma$ about the center of the spectrum in each {\em flt} image. The pixel locations of the target spectra in these images are consistent, allowing us to use the same aperture for all images except for the fifth orbit, which we shifted down by a single pixel.

The validation of our aperture photometry was meticulous. We estimated flux uncertainties using the error images packaged within the {\em flt} FITS image arrays by passing them into the {\tt photutils} aperture photometry routine. We divided the summed flux in counts and their associated uncertainties by the mean flux of the entire five-orbit light curve. Orbit by orbit, we compared the mean relative flux of the aperture photometry light curve to a light curve we generated from integrating the flux calibrated, time-tagged 1D spectra from 1800-3100\,\AA\ with the same 60-s effective exposure times and excluding pixels with problematic data quality flags. We adjusted the mean relative flux of each orbit to be in unity with that of the integrated 1D spectra light curve, the differences of which were 1-3\%, to calibrate our aperture photometry. We kept the aperture photometry light curve for our analysis because its point-to-point scatter was 4.7\%, compared to 6.2\% for the integrated 1D spectrum light curve. For a second validation, we compared our aperture photometry light curve to the light curve we generated using the software developed by the Rocky Worlds DDT program\footnote{\url{https://github.com/spacetelescope/rocky-worlds-utils}} \citep{Redfield2024}. The Rocky Worlds light curve had a higher average point-to-point scatter at 7.1\%, motivating our selection of the aperture photometry light curve for our analysis. Finally, we normalized the aperture photometry light curve in the same fashion as our ground-based light curves (Section~\ref{sec:lc_norm}).

\subsection{SOAR/Goodman Optical Spectroscopy}\label{sec:soar_spec}

We obtained a new optical spectrum of \obj on 2025~October~29 using the Goodman High Throughput Spectrograph \citep{Clemens2004} with the 4.1-m Southern Astrophysical Research Telescope (SOAR) at Cerro Pach{\'o}n, Chile, facilitated through NOIRLab AEON Program 2025B-136460. We took four 300-s exposures using the red camera with the 400 line/mm M1 grating and the 1.0\arcsec\ slit, yielding a resolving power of $R\approx1000$. We reduced these spectra and produced a final 1-D coadd using {\tt PypeIt} \citep{Prochaska2020}. We flux calibrated the spectrum using a sensitivity function in {\tt PypeIt} generated from the ESO standard star HR\,7596, which we observed with two 0.4-s exposures on the same night with the same setup used for our target. 

\begin{figure*}[!t]
    \epsscale{1.12}
\plotone{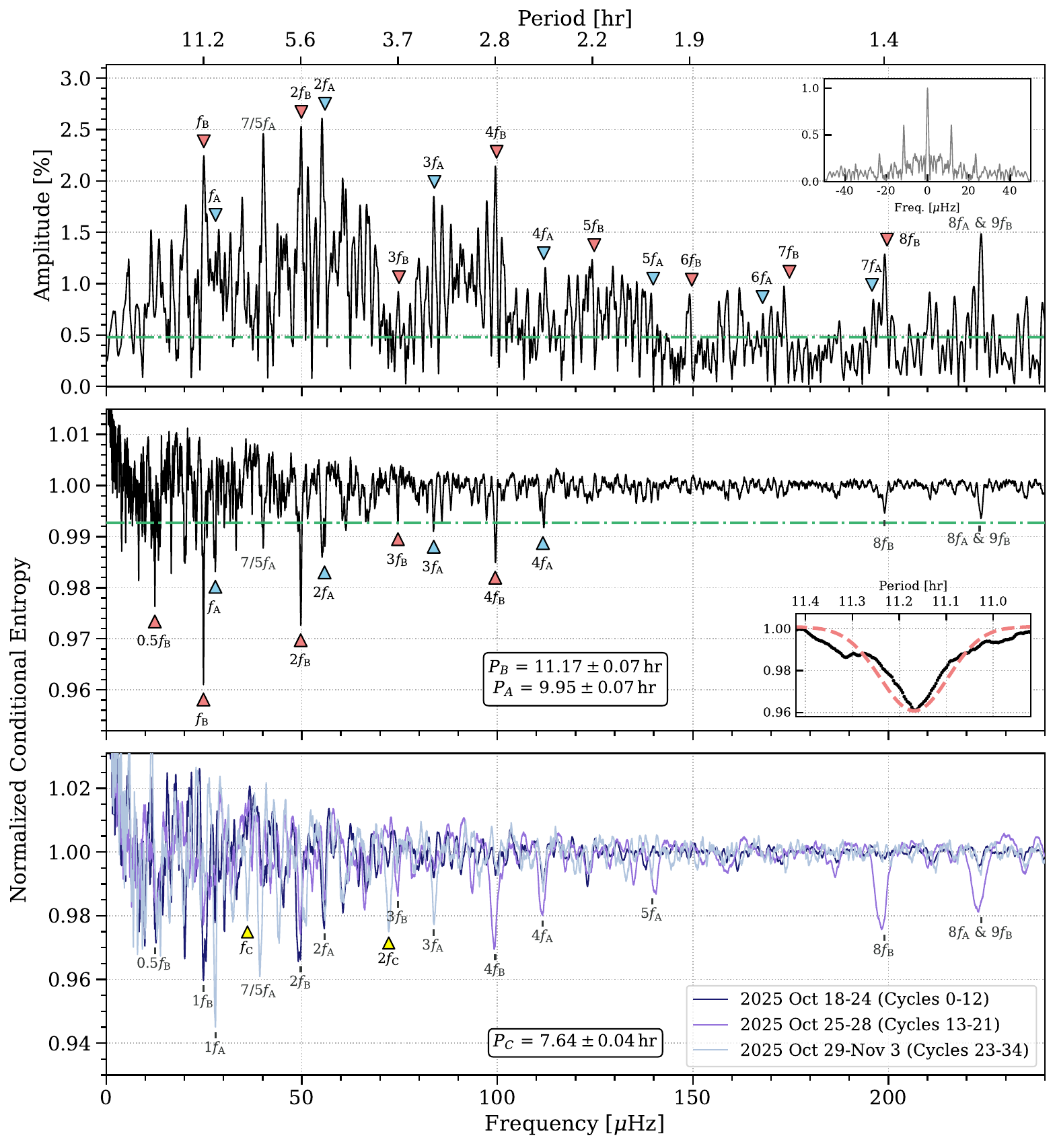}
    \caption{Lomb-Scargle (top) and conditional entropy (middle and bottom) periodograms of our 2025~autumn optical light curves of \obj. The top and middle panels incorporate the entire dataset, while the bottom panel is partitioned into three windows of $4-7$ nights. Each panel shares the same frequency axes. Multiple harmonics of $P_{\rm B}$ (red triangles) and $P_{\rm A}$ (light blue triangles) are significant beyond the bootstrapped 0.1\% false-alarm probability and 10\,$\rho$ thresholds for the Lomb-Scargle and conditional entropy periodograms, respectively, each delineated as dash-dotted green lines. In the top panel we inset the spectral window to clarify our resolution. We inset in the middle panel our Gaussian fit to the core of $P_{\rm B}$. The bottom panel highlights rapid variations in the periodicity of the transits, where over the course of the campaign the light curves shift from being dominated by $P_{\rm B}$ to $P_{\rm A}$. A new period emerges during the final window, $P_{\rm C}$ (yellow triangle). Aliases are likely present, namely 7/5$\,f_{\rm A}$. The rapidly varying periodicity suggests ongoing dynamical activity within the system that is bringing material from the various orbits into and out of view, matching the morphological complexity within the light curves of Figure~\ref{fig:Oct2025_pfold_LCs}.
    }\label{fig:periodograms}
\end{figure*}


\section{Long-Term and Rapid Variable Periodicity in the Transits}\label{sec:periodanalysis}

Our ground-based, follow-up, time-series photometry campaign was designed to begin several days in advance of our HST program so that we could readily identify and track repeating features, such as those centered around Phase\,$\approx$\,0.38 in Figure~\ref{fig:UCAM_Nov2024_pfold_LCs}. Our light curves quickly revealed the system entered a new state: transits that recurred at $P_{\rm A}$ period were no longer apparent. Transit morphologies instead aligned at $P_{\rm B}$, a period previously exclusively detected in TESS Sector~31 observations of \obj \citep{Vanderbosch2021}, and quickly lost coherence over consecutive orbits.

Here we detail our period search analysis of our ground-based, follow-up, optical photometry collected during both epochs of 2024~November and 2025~autumn. We keep only the highest fidelity light curve when overlapping coverage is present. Our period search draws on previous work that combines conditional entropy and Lomb-Scargle periodograms to resolve periodicity in transit debris \citep{Guidry2025}. We did not find improvement using the Fast Periodicity Weighting algorithm (FPW; \citealt{Finkbeiner2025}), as the forest of similarly prominent peaks in the $P_{\rm A}$ and $P_{\rm B}$ regions made period identification non-trivial at bin choices of 5, 10, and 20. FPW may struggle at reliable period detection for embedded transiting debris systems whose light curve morphologies can be rapidly dynamic. The best period values from both algorithms are reported in Table~\ref{tab:periods}.

\subsection{Conditional Entropy}\label{sec:CE}

\subsubsection{2025 October}

Conditional entropy as a period-finding algorithm is optimized for non-sinusoidally varying, irregularly-sampled time-series, being inherently agnostic to the intrinsic model driving observed variability \citep{Graham2013a, Graham2013b}, making it the ideal tool for measuring orbital periods of transiting debris \citep{Guidry2025}. 

We use GPU-accelerated Conditional Entropy ({\tt gce}\footnote{\url{https://github.com/mikekatz04/gce}}; \citealt{Katz2021}) to period search our optical light curves from 2025 October. Our search design follows precedents established by \citet{El-Badry2022, Guidry2025}. We first search the entire dataset across a frequency grid with resolution 1/$T$, where T is the total observing baseline. We oversample the grid by a factor of 100 and truncate the grid at 5000\,$\mu$Hz; no variability has been detected in the system more rapid than 1000\,$\mu$Hz \citep{Vanderbosch2021}. Our search evaluates over a grid of 30 phase bins and 10 flux bins. We normalize the resultant periodogram by subtracting a sliding median filter with a width of 7.20\,$\mu$Hz (equivalent to 1001 resolution elements), yielding the periodogram in the middle panel of Figure~\ref{fig:periodograms}. The minimum in the periodogram is located at 24.8747\,$\mu$Hz, or equivalently 11.17\,hr. We attempt to optimize this measurement by fitting a Gaussian to the core of the minimum, however there is a secondary feature at around 11.32\,hr that skews the fit. Instead, we only record the HWHM as an estimate of the uncertainty to our measurement, yielding $P_{\rm B} = 11.17 \pm 0.07\,$hr. The region around $P_{\rm A}$ is even more definitively non-Gaussian, requiring us to again simply record the local minimum of the periodogram and we adopt a 0.2$\,\mu$Hz uncertainty: $P_{\rm A} = 9.95 \pm 0.07$\,hr. Analogous to Figure~\ref{fig:Oct2025_pfold_LCs}, we illustrate our 2025~autumn light curves folded over $P_{\rm A}$ in Figure~\ref{fig:fold_A_Period}. We annotate the location of $P_{\rm B}$ and $P_{\rm A}$ and their harmonics in Figure~\ref{fig:periodograms}, several of which are significant to the nominal detection threshold ($\rho = 10 = \mu - 10\sigma$, for mean $\mu$ and standard deviation $\sigma$; \citealt{Katz2021}). The typical 0.5:1:2 harmonic pattern \citep{Graham2013a} is especially notable for $P_{\rm B}$.

To test the stability of these periods across our campaign, we partition the photometry into three windows, each a few days in length. We define trial frequency grids and phase-flux grids identically to our global search, and normalize the calculated periodograms using a  sliding median filter with a width of 1001 resolution elements (or widths of 16.9, 37.2, and 22.0$\,\mu$Hz, respectively). These periodograms are depicted in the bottom panel of Figure~\ref{fig:periodograms}. The periodogram of the 2025~October~18-24 data indicates signals from $P_{\rm B}$ and $P_{\rm A}$ and their second harmonics without higher-order structure. The periodogram takes an entirely different form over 2025~October~25-28, becoming dominated by higher-order harmonics of both periods, whose emergence suggests a more complicated transit morphology, in turn helping explain the absence of repeating patterns in Figure~\ref{fig:Oct2025_pfold_LCs} during this window. The third window of 2025~October~29-November~3 shows greatly diminished signal at $P_{\rm B}$ and a preference for $P_{\rm A}$. Like the first window, the transits are simpler with no power existing above the fourth harmonic. A new ``C'' period notably emerges within this window, $P_{\rm C} = 7.69 \pm 0.04$\,hr, measured again from the local minimum  and assuming 0.2\,$\mu$Hz uncertainty. Its second harmonic is identifiable and marked. We defer further analysis of this new period to Section~\ref{sec:c_period}.

\subsubsection{2024 November}

We also period search our 2024~November ULTRACAM photometry using {\tt gce}, adhering to the same method as with our 2025~autumn campaign. This conditional entropy periodogram is shown as the right panel of Figure~\ref{fig:UCAM_Nov2024_pfold_LCs}. We measure the minimum within the $P_{\rm A}$ region at $P_{\rm A} = 9.92 \pm 0.07$\,hr after selecting a 0.2\,$\mu$Hz uncertainty, agreeing with the original measurement by \citet{Vanderbosch2021}. We recover the best alignment when folding the light curves over 9.937\,hr as found in Figure~\ref{fig:UCAM_Nov2024_pfold_LCs}. $P_{\rm B}$ and $P_{\rm C}$ are not significant during this epoch.

\subsection{Lomb-Scargle Periodogram (2025~Autumn)}

We also compute a Lomb-Scargle periodogram \citep{Lomb1976,Scargle1982} over the same trial frequency grid used in our conditional entropy search. We employ the {\tt Astropy} implementation using PSD normalization with {\tt nifty-ls} \citep{2024ascl.soft12008G} as the evaluating method. We convert the output power, $p$, into amplitude: $A = \sqrt{|4p/N|}$, for $N$ many points in the time series. This periodogram is plotted in the top panel of Figure~\ref{fig:periodograms}, where a quick inspection reveals a complex structure with significant power from $P_{\rm B}$ and $P_{\rm A}$ and several of their harmonics, along with possible aliases. We bootstrap a 0.1\% false-alarm probability significance threshold by simulating 10${,}$000 periodograms with replacement \citep{VanderPlas2018}. The maximum frequency within the $f_{\rm B}$ region equates to $P_{\rm B} = 11.13 \pm 0.09$\,hr, where we assume an uncertainty of 0.2\,$\mu$Hz, and resonates out to at least its eighth harmonic. We detect $f_{\rm A}$ at a much weaker, albeit still significant amplitude where $P_{\rm A} = 9.93 \pm 0.07$\,hr. Several harmonics of this period also appear, with the second harmonic appearing to be merged with a higher-amplitude, unidentified peak.

We attempt to refine these periodicity measurements using {\tt Pyriod} \citep{2022ascl.soft07007B}, where we fit sinusoids of frequencies $f_{\rm B}$ and $f_{\rm A}$ to the light curves. Because linear least squares fits tend to underestimate uncertainties on fits to time series, we calculate uncertainties in the frequency, amplitude, and phase following \citet{Montgomery1999}. We find best-fit values of $P_{\rm B} = 11.20 \pm 0.15$\,hr and $P_{\rm A} = 10.06 \pm 0.11$\,hr, both in agreement with our conditional entropy measurements to at least $1\sigma$. The less accurate $P_{\rm A}$ measurement may be a consequence of the nearby beating or aliasing and relatively weaker signal.

\begin{deluxetable}{cccc}[!t]
\tablenum{1}
\tablecaption{Period Determinations from Conditional Entropy and Lomb-Scargle Periodograms\label{tab:periods}}
\tabletypesize{\footnotesize}
\tablehead{
    \colhead{Method} & \colhead{$P_{\rm A}$} & \colhead{$P_{\rm B}$} & \colhead{$P_{\rm C}$}\\
    \colhead{} & \colhead{hr} & \colhead{hr} & \colhead{hr}}
\startdata
2024~Nov & & & \\
CE           & $9.92 \pm 0.07$ & - & -\\
\hline
2025~Oct/Nov &  &  & \\
CE           & $9.95 \pm 0.07$ & $11.17 \pm 0.07$ & $7.69 \pm 0.04^\dagger$\\
LS           & $9.93 \pm 0.07$ & $11.13 \pm 0.09$ & $7.65 \pm 0.04$ \\
{\tt Pyriod} & $10.06 \pm 0.11$ & $11.20 \pm 0.15$ & - \\
\enddata
\tablecomments{For each period, conditional entropy (CE) measurements are taken directly from the minima of the cores of each period. The Lomb-Scargle (LS) measurements are likewise read directly from the local maxima. We assume 0.2\,$\mu$Hz uncertainty, except for the CE measurement of $P_{\rm B}$ during 2025~October and November, which we were able to fit with a Gaussian and record the HWHM as a conservative error. We also show our measurements derived using {\tt Pyriod} with uncertainties derived using the methods from \citet{Montgomery1999}. Hyphens denote non-detections. $\dagger$: We refine this measurement by eye, observing the transits best phase at $P_{\rm C}=7.64$\,hr (Figure~\ref{fig:fold_C_Period}).}
\end{deluxetable}

\subsection{7.64-hr Drifter in 2025~October \& November}\label{sec:c_period}

We mark in the bottom panel of Figure~\ref{fig:periodograms} a potential new period that emerges during the final few days of our 2025~autumn observing season, $P_{\rm C} = 7.69 \pm 0.04$\,hr. We are able to map this period to a recurring transit captured by our RVO light curves during the consecutive nights of 2025~October~29-November~1. Folding these light curves over this period, however, does not align the $\approx$10\% deep transit in phase. Instead, we find $P_{\rm C}=7.64$\,hr best reproduces the periodicity, which we plot in Figure~\ref{fig:fold_C_Period}. We formally adopt $P_{\rm C}=7.64\pm0.04$\,hr as the measurement of this new period. It is unclear how stable this period is and its formal significance in the periodogram is less straightforward to assess. Our November~2-3 RVO light curves lack the phase coverage needed to observe the transit recur. The emergence of this  new transit period evokes the ``drifters'' at WD\,1054$-$226, which were repeating transits found on orbits clearly distinct from the 25.02-hr fundamental orbital period \citep{Farihi2022}.

\subsection{An Alias of $P_{\rm A}$ at Approximately 7\,hr}

We briefly note that during both the 2024~November and 2025~autumn observing seasons we detect a significant period of about 7\,hr in our periodograms. Conditional entropy indicates a period of 6.96\,hr within our 2024~November ULTRACAM light curves. We recover power sustained near this period in both our Lomb-Scargle and conditional entropy periodograms of the 2025~autumn light curves at 6.92\,hr. A sharp minimum emerges at 7.07\,hr during the third window of our observations (2025~October~29-November~3), traced by the pale blue conditional entropy periodogram in the bottom panel of Figure~\ref{fig:periodograms}.

We infer this period is an alias of $P_{\rm A}$. Our ULTRACAM light curves sample seven cycles of $P_{\rm A}$. If this period is the second harmonic of a 13.92\,hr modulation, our light curves sample five cycles. When folded over 13.92\,hr, we see the deep transit align in phase at cycles 0 and 5. Indeed, $7 \times 9.937 = 5 \times(2 \times 6.96) = 69.6$. We read the presence of this alias within our 2025~autumn light curves as additional evidence for the persistence of $P_{\rm A}$. We mark on Figure~\ref{fig:periodograms} the locations of this alias at $7/5 f_{\rm A}$, which appears distinct and unrelated to the nearby $P_{\rm c}$.

Other aliases are likely present in our periodograms. For example, peaks are present at both roughly 33.3 and 61.2\,$\mu$Hz (equivalently 8.35 and 4.54\,hr) appear significant, yet we observe no discernible repeating patterns in the light curves folded over these periods. The same is true for the peak at roughly 44\,$\mu$Hz seen in the October~29-November~3 window. We do not exhaustively test for variability at every nominally significant period, only the most prominent.

\section{Transit Color Term Analysis}\label{sec:color}

Previous attempts to measure a color term to the transits at \obj found the $g$- and $r$/$R$-band transit depths consistent with each other, both using the public ZTF DR5 light curve \citep{Vanderbosch2021} and follow-up photometry from 2023~December--2024~January \citep{Gary2024}. We update this exploration using our simultaneous multi-color ULTRACAM photometry from 2024~November, when transits recurred at $P_{\rm A}$, and also with our near-concurrent NUV and optical light curves from 2025~October, when $P_{\rm B}$ was predominant along the line of sight. We explore linear transit color terms of the form $y= m(\lambda) \cdot x + b(\lambda)$, where the slope corresponds $m(\lambda) = E(\lambda_b - \lambda_r) / A_r$ for given blue ($b$) and red ($r$) bandpasses.

\begin{figure*}[!t]
    \epsscale{1.15}
    \plotone{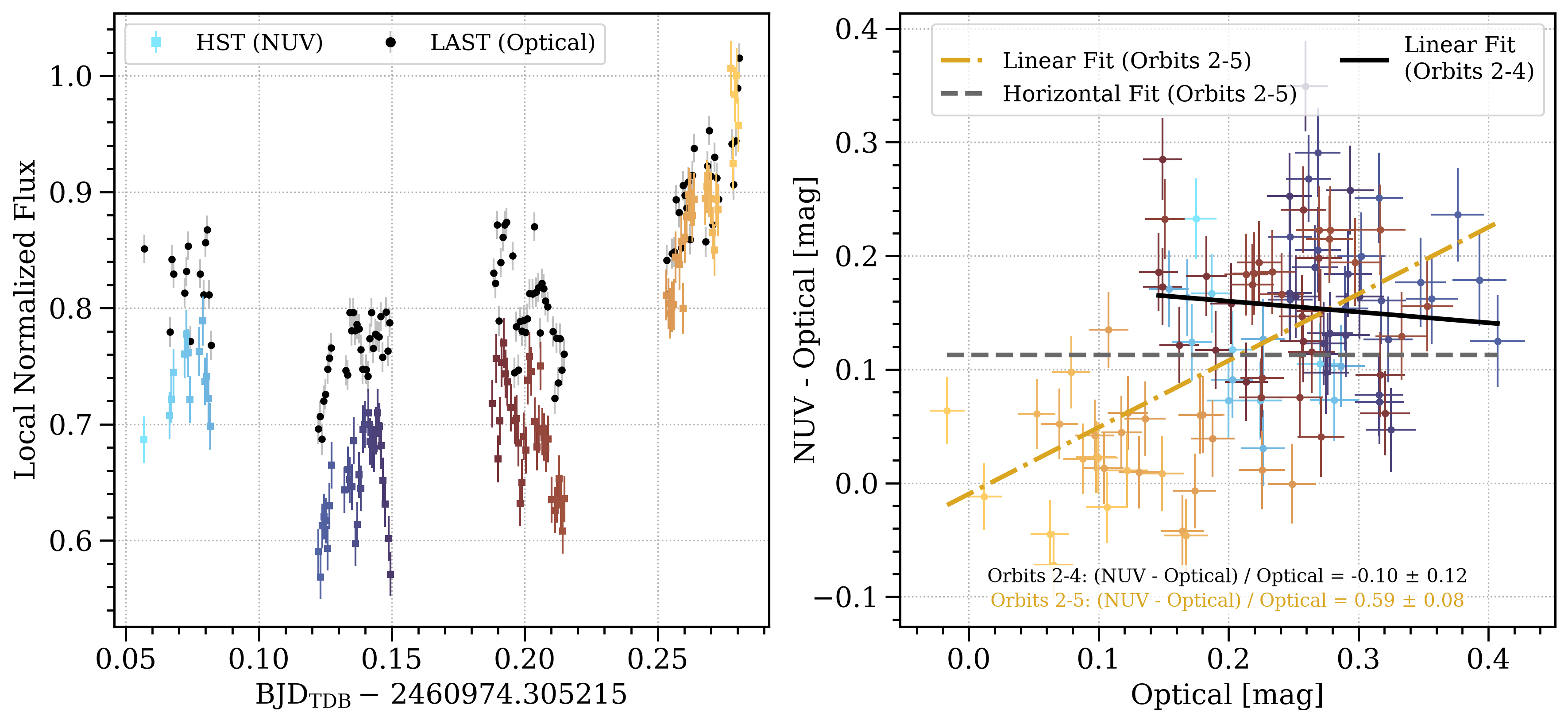}
    \caption{Comparison between the transit depths as seen in the NUV with HST/STIS to the optical as observed by LAST (binned from 30-s to 60-s exposures to match the effective HST sampling). We calculate the transit color term in the right panel using only data separated by no more than 60\,s. We use orthogonal distance regression to measure the reddening law $E(\mathrm{\textsc{nuv}} - \mathrm{\textsc{optical}}) / A_\mathrm{\textsc{optical}} = 0.59 \pm 0.08$. The residual variance of the best-fit line (4.06; solid gold line) is preferred to the null hypothesis, the fixed horizontal line (5.84; dashed gray line). This color term is intrinsic to the data during fifth HST orbit; we measure $E(\mathrm{\textsc{nuv}} - \mathrm{\textsc{optical}}) / A_\mathrm{\textsc{optical}} = -0.10 \pm 0.12$ considering only the second through fourth HST orbits. The particle size distribution could significantly vary dip-to-dip and over consecutive orbital cycles as collisions rearrange the transiting debris, however, the fifth orbit data are the most normalization-dominated and their transits the shallowest. Over HST orbits 2-4, NUV starlight is constantly preferentially extincted given by the best-fit y-axis intercept of the linear fit being $A_{\mathrm{\textsc{nuv}}} = A_{\mathrm{\textsc{optical}}} + 0.18\,$mag.
    }\label{fig:HST_LAST_Colors}
\end{figure*}

\subsection{Colorless Transits along $P_{\rm A}$ in 2024~November}\label{sec:ucam_colors}

We measure the transit depths within our simultaneous $ugr$ ULTRACAM photometry to be slightly inconsistent across the wavelengths sampled. We emulate previous experiments using our simultaneous $ugr$ ULTRACAM photometry to search for linear trends in the $E(u-g)/A_g$, $E(g-r)/A_r$, and $E(u-r)/A_r$ color distributions \citep{Farihi2022,Guidry2025}. We use orthogonal distance regression \citep{Boggs&Donaldson1989} as implemented by {\tt scipy} \citep{2020SciPy-NMeth} to determine the extinction laws between these various blue and red bands. The best-fit color terms are as follows: $E(u-g)/A_g = 0.022 \pm 0.006$, $E(g-r)/A_r = -0.088 \pm 0.006$ and $E(u-r)/A_r = 0.007 \pm 0.007$. Since only the $E(u-r)/A_r$ law is consistent with no color term, we evaluate the significance of the possible reddening from $u$ to $g$ and the possible bluing from $g$ to $r$ by running an MCMC simulation developed by \citet{odrfitter}, where we recover the same color terms with uncertainties that agree out to three decimal places. We further compare each distribution against the null hypothesis of colorless transits, modeled by a horizontal line with a fixed slope of zero, finding each distribution is consistent with a zero-slope trend. The residual variances of the best-fit and zero-slope lines are both 0.81 for the $u$ vs. $g$ law,  0.86 and 0.90, respectively, for the $g$ vs. $r$ law, and both 0.85 for the $u$ vs. $r$ law. Despite the marginal preference for the bluing from $g$ to $r$, we cannot claim a detection of any color term over colorless transits with rigid statistical significance. 

Without constraints on the mineralogy we can only make tenuous estimates on the particle size distributions of the transiting debris from these ULTRACAM benchmarks. Dust grains that extinguish starlight at equal efficiencies with characteristic sizes $s$ satisfy $X \equiv \frac{2 \pi s}{\lambda} \gtrsim 2$ \citep{Xu2018}. The limits from our $ugr$ ULTRACAM photometry then require $s \gtrsim 0.1\,\mu$m. The distribution of local ISM dust grains peaks at an effective size of $s \sim 0.2\,\mu$m \citep{Mathis1977,Kim1994,Kim1995}, and characteristically extincts starlight by $E(u-g)/A_g = 0.315 \pm 0.014$ and $E(g-r)/A_g = 0.287 \pm 0.005$ \citep{Wang2019,Farihi2022}. Dust-less transits by macroscopic debris can be produced via the YORP effect  \citep{Veras2020}, or tidally disrupted rubble pile debris that has yet to be collisionally ground-down into smaller grains \citep{McDonald2021,Steckloff2026}. Alternatively, the colorless transits we observe may simply reflect the optical depth of the transiting debris. Irrespective of the grain size distribution, disrupted debris that is optically thick will cast transits with no chromaticity \citep{Izquierdo2018}.

\subsection{Possibly Variable Color Terms between the NUV and Optical in 2025~October}

Inspired by a previous experiment on WD\,1145+017 \citep{Xu2019a}, in Figure~\ref{fig:HST_LAST_Colors} we compare the difference in the transit depths between our HST/STIS light curve in the NUV to the near-concurrent optical follow-up obtained with LAST. For this measurement we consider only observations occurring during the second, third, and fourth HST orbits (no LAST data overlap with the first HST orbit); observations during the fifth HST orbit are the most normalization and noise dominated. We define \textsc{nuv} and \textsc{optical} bands by convolving our HST/STIS spectrum and the LAST QE profile through our model spectrum (Section~\ref{sec:model_spec}), yielding effective wavelengths of $\lambda_{\rm eff,~NUV} = 2672\,$\AA\ and $\lambda_{\rm eff,~optical} = 5498\,$\AA.

We obtain the NUV color term measurement using the same orthogonal distance regression method we applied to our ULTRACAM photometry, pairing the nearest HST and LAST observations taken within 60\,s, equivalently one exposure time (we binned the LAST light curve by a factor of two from 30-s exposures into 60-s effective exposures). A total of 79 HST epochs are considered, separated on average 17\,s from the nearest LAST timestamp. There are no repeat observations within the 79 epochs, which span an observing baseline of 3.8\,hr. We detect no significant color term between the NUV and optical transits: $E(\mathrm{\textsc{nuv}} - \mathrm{\textsc{optical}}) / A_{\mathrm{\textsc{optical}}} = -0.10 \pm 0.12$. This is corroborated by the residual variance of the best-fit line (2.98) being consistent with that of the zero-slope, horizontal line (2.96). We validated the uncertainty on this result by performing the same MCMC analysis as we did with our ULTRACAM color terms, finding no difference. While there is no significant trend in color with transit depth, the best-fit y-intercept over HST orbits 2-4 indicates a constant color offset of $A_{\mathrm{\textsc{nuv}}} = A_{\mathrm{\textsc{optical}}} + 0.18\,$mag, quantifying the relative extinction of NUV starlight during this epoch.

A significant reddening term emerges if we include the LAST photometry overlapping with the fifth HST orbit: $E(\mathrm{\textsc{nuv}} - \mathrm{\textsc{optical}}) / A_{\mathrm{\textsc{optical}}} = 0.59 \pm 0.08$. With this addition of 28 epochs, the total baseline spanned lengthens to roughly 5.4\,hr. This significant reddening is confidently preferred to the null hypothesis of a horizontal, zero-slope distribution, with respective residual variances of 4.06 and 5.84. At the same time, we raise the caveat that both the NUV and optical transits during the fifth HST orbit are the shallowest and normalization dominated, and thus the most noise dominated. The pointing of HST was reliable for each of the five orbits and without anomalies (Section~\ref{sec:hubble_obs}). Notwithstanding this potential limitation, it is still plausible that the transits in view during the fifth orbit are drawn from a different distribution of particle sizes that do not extinct NUV light as efficiently relative to the transits of the second, third, and fourth orbits.

While deep, long-period transits have reddened white dwarfs \citep{Bhattacharjee2025a,VanRoestel2026}, similar experiments measured the transits at WD\,1145+017 \citep{Alonso2016,Zhou2016,Croll2017,Xu2018}, WD\,1054$-$226 \citep{Farihi2022}, and ZTF\,J1944+4557 \citep{Guidry2025} all to be wavelength-independent in the optical\footnote{The in-transit UV bluing of WD\,1145+017 in the ultraviolet can be explained by diminished absorption from circumstellar gas during transits \citep{Hallakoun2017,Xu2019a}.}.

\section{Updated NUV-Optical Model Spectrum}\label{sec:model_spec}

\begin{figure}[!t]
    \epsscale{1.15}
    \plotone{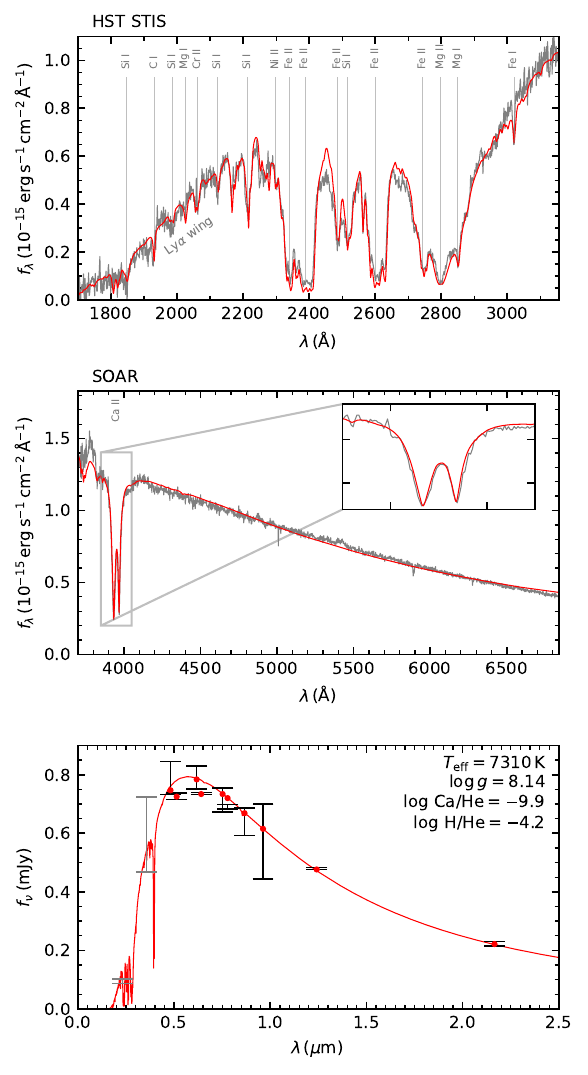}
    \caption{Model spectrum of \obj (red) superimposed over our HST/STIS spectrum in the NUV (top, gray) and optical SOAR/Goodman spectrum (middle, gray). We highlight several prominent metal transitions seen as absorption and the Lyman $\alpha$ wing in the NUV, along with insetting a magnification of the optical Ca-II H and K transitions. Our SOAR spectrum exhibits absorption at the sodium D doublet ($\lambda \approx 5900\,$\AA) with no detectable absorption at H$\alpha$. The bottom panel compares  the SED photometry to the model from the NUV to the NIR. Red circles correspond to the model fluxes and the photometry is shown as the black error bars without markers. The silver error bars represent the SDSS $u$ and GALEX {\sc nuv} photometry and are only shown for reference. 
    }\label{fig:model_spec}
\end{figure}

We update the model spectrum of \obj published by \citet{Vanderbosch2021} through joining our HST/STIS NUV spectrum with a new, nearly concurrent SOAR/Goodman optical spectrum (Figure~\ref{fig:model_spec}). We imitate their process by using the same \citet{Coutu2019} fitting procedure with the \citet{Blouin2018} DZ model atmospheres. Given flux excursions of ${\gtrsim} 10$\% are consistently observed toward \obj (Figure~\ref{fig:UCAM_Nov2024_pfold_LCs}), we suspect the catalog photometry of \obj is not reliable for flux calibration. Instead of anchoring our model to the published Pan-STARRS DR2 photometry ($\sigma_r = 0.006\,$mag; \citealt{2020ApJS..251....6M,2025yCat.2389....0M}), we generate synthetic Pan-STARRS photometry by convolving the Gaia XP spectrum \citep{Carrasco2021,DeAngeli2023} through the published response curves to calibrate our estimate of the photosphere \citep[see][]{Arseneau2026}. Gaia XP spectra (which span $3300-10{,}500$\,\AA\ with $R\approx50$) are the average of every BP and RP transit \citep{DeAngeli2023}; \obj has at least 36 in each channel in Gaia DR3. The resultant synthetic photometry are inherently more uncertain but are likely more representative of the out-of-transit flux baseline for the white dwarf. Unlike \citet{Vanderbosch2021}, we elect to incorporate the near-infrared $J$ and $K_S$ VHS photometry into our fitting process to replace our highly uncertain synthetic $z$-band benchmark. We exclude the unWISE photometry from our fit, despite being consistent with the white dwarf photosphere, since their inclusion worsens the fit to the STIS spectrum and does not significantly affect the derived atmospheric parameters ($\Delta T_{\rm eff} = 80\,$K, for example). We report our best-determined atmospheric parameters in Table~\ref{tab:atmos_params}.

\begin{deluxetable}{lcc}[!b]
\tablecaption{Model NUV-Optical Spectrum Atmospheric Parameters of \obj}
\label{tab:atmos_params}
\tablewidth{0pt}
\tablehead{\colhead{Parameter} & \colhead{Value} & \colhead{Uncertainty}} 
\startdata
$T_{\mathrm{eff}}$ (K)      & 7310      & 100   \\
$\log(g)$ (cgs)             & 8.14      & 0.03 \\
$M_{\rm wd}$ (M$_{\odot}$)   & 0.662     & 0.020 \\
$R_{\rm wd}$ (R$_{\odot}$)   & 0.0107    & 0.0002\\
$[$H/He$]$                  & $-$4.20   & 0.20  \\
$[$C/He$]$                  & $-$8.70   & 0.30  \\
$[$O/He$]$                  & ${<}-5$   &       \\
$[$Mg/He$]$                 & $-$8.70   & 0.12  \\
$[$Si/He$]$                 & $-$9.01   & 0.14  \\
$[$S/He$]$                  & ${<}-5$   &       \\
$[$Ca/He$]$                 & $-$9.89   & 0.12  \\
$[$Cr/He$]$                 & $-$10.31  & 0.27  \\
$[$Fe/He$]$                 & $-$8.56   & 0.13  \\
$[$Ni/He$]$                 & $-$9.61   & 0.30  \\
$[$Mg/Si$]$                 & 0.31      & 0.18  \\
$\tau_\mathrm{cool}$ (Gyr)  & 1.77      & $^{+0.12}_{-0.10}$  \\
$\tau_\mathrm{total}$ (Gyr) & 3.20      & $^{+1.40}_{-0.50}$  \\
\enddata
\tablecomments{$\tau_\mathrm{cool}$ and $\tau_\mathrm{total}$ are estimated by {\tt wdwarfdate} \citep{Kiman2022} using a non-DA model IFMR \citep{Cummings2018}.}
\end{deluxetable}

\begin{figure}[!t]
    \epsscale{1.15}
    \plotone{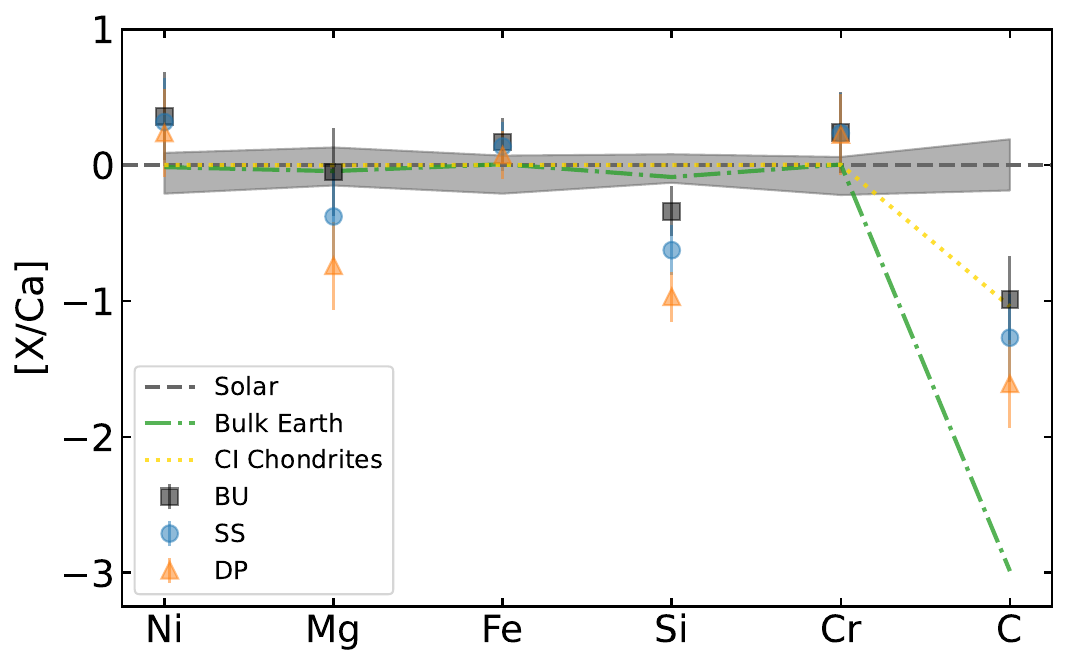}
    \caption{Photospheric abundance ratios of select elements with respect to calcium derived from our model spectrum of \obj, all normalized to the Solar metallicity (gray dashed line). We consider build-up (charcoal squares), steady state (blue circles), and declining phase (orange triangles) accretion scenarios. The debris is consistent to within 1\,$\sigma$ with a composition akin to CI chondrites (yellow dashed-dotted line) and delivered via build-up accretion, except for silicon. Our carbon detection is about 2\,dex enriched relative to the expectation for bulk Earth material (green dashed-dotted line).
    }\label{fig:abundances}
\end{figure}

We detect several transitions from multiple elements, mostly from Fe~II and Mg~II. The C~I line at 1930\,\AA\ stands out in particular. We estimate [C/Ca]\,$= 1.19 \pm 0.32$, consistent with a chondritic-like body ($[$C/Ca$]=1.14$ for CI chondrites \citealt{Lodders2025}). The evidence for a chondritic composition is illustrated in Figure~\ref{fig:abundances}, where we observe the abundance ratios relative to the Solar metallicity of the accreted debris to be consistent with chondrites to within 1$\sigma$ at all elements except silicon using models from \citet{Dufour2012} \citep[see][]{Rogers2025b}. Both build-up (BU) and steady state (SS) accretion scenarios can explain the derived abundances ratios; declining phase (DP) accretion is much less likely. While the nickel, magnesium, iron, silicon, and chromium abundance ratios are also consistent with that of bulk Earth, our carbon detection is appreciably discrepant. Convective dredge-up should uplift some carbon to the photosphere of \obj \citep{Bedard2022,Blouin2023b}. A $T_{\rm eff} = 7300$\,K and $M_{\rm wd} = 0.65\,$M$_{\odot}$ hydrogen-deficient white dwarf is expected to show $[$C/He$] = -10.4 \pm 1.2$ due to dredge up in the PG\,1159 progenitor \citep{Blouin2023}. Our detection of $[$C/He$]=-8.7 \pm 0.3$ suggests a significant fraction of the observed carbon is accreted onto \obj. We do not detect any other volatile species besides carbon, although we estimate upper limits of $[$O/He$] < -5$ and $[$S/He$] < -5$ from our model.

At $[$Mg/Si$] = 0.31 \pm 0.18$, our detection appears to be the highest yet observed for a polluted white dwarf with characterized disk compositions, though higher $[$Mg/Si$]$ values are known in the broader population \citep{Williams2024}. A high $[$Mg/Si$]$ ratio may indicate disk silicates dominated by olivine (Mg$_2$SiO$_4$) rather than pyroxene (MgSiO$_3$) \citep{Rogers2025b}, assuming olivine is in the Mg-rich form, forsterite. Although overlapping accretion from compositionally distinct parent bodies can produce a photospheric $[$Mg/Si$]$ ratio that does not reflect any single parent body, the high value observed here makes this less likely.

The red Lyman-$\alpha$ wing dominates the bluest portion of our NUV spectrum, granting us the ability to estimate the photospheric hydrogen-to-helium abundance ratio, and in turn grounding all other abundance ratios we report. Our derived $[$H/He$]=-4.2 \pm 0.2$ is consistent with what is observed in field DZs \citep{Hollands2017,Coutu2019}, and can be explained by the convective mixing of a superficial hydrogen layer with the helium envelope underneath. At $T_{\rm eff} = 7300$\,K, based on Table~3 of \citet{Rolland2018} and assuming a 0.6\,M$_{\odot}$ white dwarf, mixing would have occurred around $T_{\rm eff} = 9000-10{,}000$\,K, and when the spectral type would have been DA(Z) and the hydrogen layer mass would be of order $10^{-10}-10^{-9}\,$M$_{\odot}$. At this current epoch, our $[$H/He$]=-4.2$ detection supports $\log(M_{\rm H}/M_{\rm WD}) \simeq -10$ \citep{Rolland2018}.

\obj has accreted a fraction of calcium ($[$Ca/He$] = -9.89 \pm 0.12$) roughly 1\,dex lower than DZs with similar effective temperature \citep{Dufour2007,Hollands2017}. Extrapolating from our hydrogen layer mass benchmark $\log(M_{\rm H}/M_{\rm WD}) = -10$, we estimate $M_{\rm Ca} \simeq 10^{19}$\,g. We cannot confidently rule what amount of the trace hydrogen in the photosphere is accreted. With this much calcium deposited onto the photosphere, \obj does not appear to be an extreme outlier like some metal-polluted white dwarfs that appear to have excess hydrogen relative to the calcium they have accreted \citep{GentileFusillo2017}, which is a signpost for the possible accretion of a water-enriched body.

All detected metal lines in our HST/STIS spectrum appear consistent with a photospheric origin at this resolution. Our model extends into the optical in the middle panel of Figure~\ref{fig:model_spec}, where the Na-I~D doublet ($\lambda_{\rm D} \approx 5890\,$\AA) detected in our SOAR/Goodman is notably not reproduced. The high-resolution MIKE spectrum ($R \approx 22{,}000$) collected by \citet{Vanderbosch2021} captured these sodium transitions as circumstellar gas in absorption, given their distinct velocities relative to the weak H-$\alpha$ line they detected (we do not find absorption at H-$\alpha$ in our SOAR/Goodman spectrum). While our model appears to account for all detected transitions in the NUV assuming only photospheric absorption, it is possible some of the absorption we observe may be due to unresolved circumstellar gas. If present, unaccounted gas would skew the accuracy of our derived abundances \citep{LeBourdais2024}.

The atmospheric parameters of our updated joint NUV-optical model spectrum are largely consistent with what was previously estimated by \citet{Vanderbosch2021}. Notably, the white dwarf mass decreases by about $0.1$\,M$_{\odot}$. Using Gaia photometry alone, \citet{GentileFusillo2021} estimate $M_{\rm wd} = 0.876 \pm 0.039$\,M$_{\odot}$ assuming a helium-dominated atmosphere. The inclusion of our HST/STIS spectrum tightens our constraint on what the true, out-of-transit, photospheric flux is, compared to the underluminous, transit-dominated Gaia estimate. Our model also agrees with the atmospheric parameters derived by \citet{Kilic2026} from their fit of the publicly cataloged SED photometry and the DESI spectrum of \obj ($T_{\rm eff} = 7392 \pm 66\,$K, $\log(g) = 8.161 \pm 0.013$, and $M = 0.676 \pm 0.011\,{\rm M}_{\odot}$) and with the bespoke DZ modeling done by \citet{OuldRouis2026b} ($T_{\rm eff} = 7380 \pm 70\,$K, $\log(g) = 8.16 \pm 0.03$, and $M = 0.672 \pm 0.011\,{\rm M}_{\odot}$).

\section{Discussion}\label{sec:discussion}

Our campaigns of follow-up, time-series photometry and spectroscopy of \obj establish a few key results:

\begin{itemize}
    \item $P_{\rm A} = 9.937$\,hr has remained stable through our 2024~November and 2025~autumn observing campaigns, consistent with the original discovery light curves collected in 2020~December and 2021~February and TESS Sectors 4 and 31 light curves analyzed by \citet{Vanderbosch2021}.
    \item Our light curves in 2025~October and November show rapidly variable periodicity; the line of sight evolves from being dominated by transits recurring on $P_{\rm B} = 11.17$\,hr to $P_{\rm A}$-dominant. In particular, the prominence and harmonic structure of these periods in the conditional entropy periodogram varied over the roughly two weeks we observed the system, correlating with the heightened morphological complexity we observe in our light curves. Additionally, a new candidate period emerges within October~29-November~1 RVO light curves at $P_{\rm C}=7.64$\,hr.
    \item The color of the transits does not vary with transit depth. The color terms we measure using our simultaneous $ugr$ ULTRACAM photometry from 2024~November are consistent with being wavelength independent. Near-concurrent NUV and optical time-series photometry from 2025~October when the white dwarf was most extremely occulted demonstrates a preferential extinction of NUV flux that is constant with transit depth. A statistically significant reddening term emerges during the fifth HST orbit, but arrives with the caveat that these particular data are the most normalization dominated and include the shallowest transits.
    \item Our joint NUV-optical spectrum collected in late 2025~October updates the census of accreted elements in the photosphere, which includes carbon. Our abundance analysis supports that the white dwarf is accreting debris consistent with originating from a chondritic-like parent body.
\end{itemize}

\subsection{Collisions Can Explain Variable Periodicity}

Never before has a change in the predominant orbital period of a transiting debris system been so clearly observed. Orbital periods have been unreliable for predicting transits, namely at ZTF\,J0139+5245 \citep{Vanderbosch2020} and SBSS\,1232+563 \citep{Hermes2025}, and periodic transits have vanished at WD\,1145+017 \citep{Aungwerojwit2024} and ZTF\,J1944+4557 \citep{Guidry2025}, eventually resuming at the latter. $P_{\rm B}$ at \obj was previously only known to be transient, appearing exclusively during the first half of TESS Sector~31 \citep{Vanderbosch2021}.

Transience in white dwarf debris disks is most easily observed as photometric variability in the infrared. The collisional cascade in the disk surrounding WD\,0145+234 inflated the emitting area enough to brighten the system by a factor of about 2.5 at both 3.4 \& 4.6\,$\mu$m \citep{Wang2019_WD0145,Swan2021}. The system returned to its pre-collisional cascade state after about 3\,yr \citep{Swan2024}. Collisions at WD\,J210034.65+212256.89 generated a significant brightening too, especially in the $K_{\rm S}$ band, with correlated intensification of the gaseous emission at the calcium triplet in particular \citep{Rogers2025a}. Gas emission correlates with dramatic near-infrared variability across the class of dusty white dwarfs \citep{Swan2020,Guidry2024,Noor2025,Noor2026}, underscoring collisions as a critical production site for observed circumstellar gas.

Minor collisions are likely regularly occurring in the disk around \obj, given the continuously evolving transit morphologies observed in Figures~\ref{fig:UCAM_Nov2024_pfold_LCs} and \ref{fig:Oct2025_pfold_LCs}, and also light curves of the system published in the literature \citep{Vanderbosch2021,Aungwerojwit2024,Korth2026}. The circumstellar gas in absorption observed by \citet{Vanderbosch2021} may further lend credence to these ongoing collisions. We postulate that transits associated with $P_{\rm B}$ could have become the predominant pattern along the line of sight if a collisional cascade occurred within this orbit, boosting its occulting fraction enough to overtake that of the  debris transiting on the $P_{\rm A}$ orbit, either through elevating the optical depth or increasing the ring's scale height.

Collisions are likely necessary to grind down disrupted fragments into the observed dust in white dwarf debris disks \citep{Wyatt2011,Swan2020,Li2021,Brouwers2022}. For example, rubble piles orbiting white dwarfs at periods $P \gtrsim 10\,$hr cannot be ground down into particles smaller than $10^2-10^3$\,m without collisions \citep{Steckloff2026}. Fragments along eccentric orbits, like those associated with $P_{\rm A}$ and $P_{\rm B}$, inevitably collide and eventually grind down into sub-micron grains \citep{KenyonBromley2017a,KenyonBromley2017b,Malamud2020a,Malamud2021}. While it appears the transit depths in 2025~October were most likely colorless as a function of transit depth, the possible in-transit reddening during the fifth orbit of our HST program commensurate with the transits beginning to recur at $P_{\rm B}$ could be evidence for recent dust production by collisions within the B ring. However, we are unable to determine whether the material that cast this possible preferential extinction of NUV light is associated with the A or B orbit as the transits superimpose over each other.

Disks are predicted \citep{Dominik2003,KenyonBromley2017a,KenyonBromley2017b} and observed \citep{Swan2024} to return to pre-cascade states following collisions.  $P_{\rm A}$ has been the predominant if not only period observed within the system during all previous epochs except 2025~autumn. If a collisional cascade occurred within the B orbit, the system should eventually return to being dominated by transits repeating at $P_{\rm A}$; we may have observed this settling as $P_{\rm A}$ returned as the predominant period during the third and final window of 2025~autumn season (Figure~\ref{fig:periodograms}). Direct evidence for a collisional cascade would be found in an infrared brightening circa 2025~October, similar to what was observed at WD\,0145+234 and WD\,J210034.65+212256.89.

Instead of collisions there could be a cyclical or periodic process that brings the B ring into and out of view independent of collisions. Gas disks around white dwarfs are expected to general relativistically precess over years-long timescales \citep{Miranda2018,Goksu2024}, and the radial velocities of gas disk emission profiles are observed to vary over these timescales \citep{Manser2016,Manser2021}---except for the especially rapid variability at HE\,1349$-$2305 \citep{Dennihy2018}---as does the gas disk viewed in absorption around WD\,1145+017 \citep{Cauley2018}. Precession was posited as a possible explanation to the approximately 2\,yr-long transit of ZTF\,1013$-$0427 \citep{Bhattacharjee2025a}, another white dwarf with a gas disk in emission, which significantly reddened the star as observed by ZTF in the $g$ and $r$ bands. The transits at ZTF\,J1944+4557 appear to oscillate from being in view and out of view every 1.5-2\,yr \citep{Guidry2025}. Long-term, multi-year monitoring will be necessary to rule out the possibility that a cyclical or periodic process is responsible for the emergence of transits repeating at $P_{\rm B}$ by instead observing these transits arise sporadically, possibly akin to SBSS\,1232+563 \citep{Hermes2025}. Including our observations, transits repeating at $P_{\rm B}$ have never been observed to persist beyond two weeks, although the time of their emergence for both epochs is unconstrained.

Another possibility may be that various independent orbits destabilized one another. Transit light curves of WD\,1145+017 suggest several co-orbital rings of debris at distinct periods \citep{Gaensicke2016}, a model reproduced by $N$-body simulations \citep{Veras2016}. Co-orbital debris rings are subject to dynamical instabilities on timescales of $t_{\rm inst} \sim 2$\,yr if the orbits are not nearly circular \citep{Gurri2017}. Since each of $P_{\rm A}$, $P_{\rm B}$, $P_{\rm C}$ require eccentric orbits to disrupt an asteroid analog \citep{Rappaport2013,Veras2014,Veras2017,Vanderbosch2021}, it is plausible the erratic, evolving transit morphologies and the alteration of the predominant period may reflect a destabilization of the system. The bottom panel of Figure~\ref{fig:periodograms} spotlights the incoherence of the periods in the system over our observing baseline in 2025~autumn, unlike in 2024~November where the transit morphologies of the A period appeared to be more consistent orbit-to-orbit (Figure~\ref{fig:UCAM_Nov2024_pfold_LCs}). Previous simulations demonstrated interactions with an external body of mass $M_{\rm external}=10^{23}$\,g are capable of perturbing the A and B orbits at \obj \citep{Vanderbosch2021}.

\subsection{An Emerging Dichotomy of Transiting Debris Light Curve Morphology}\label{sec:morphology}

Transit light curve morphologies thus far appear categorizable into two modes: (1) ``discrete'' and (2) ``relentless'' or ``embedded.'' Systems such as ZTF\,J1944+4557, on the short-period end, and ZTF\,J0923+4236 and ZTF\,J0139+5245, at the long-period end, have shown phases of clearly defined, discrete phases of starlight un-occulted by transiting debris. On the other hand, ZTF\,J0328$-$1219, WD\,1054$-$226 \citep{Farihi2022}, and ZTF\,J0923+7326 \citep{Bhattacharjee2025a} are always observed to be in transit. These stars appear embedded behind their disks which continuously block flux as they revolve, and are thus constantly underluminous. It is possible that embedded transiting systems either host disks with the largest vertical heights and/or are the most edge-on among the class; the degree to which they are slightly inclined off of 90\,deg should have observable consequences \citep{Bhattacharjee2025b}.

Transit morphologies, irrespective of their discrete or embedded nature, are dynamic. The activity levels of WD\,1145+017 have been regularly and densely monitored since 2015, showing phases of heightened and diminished activity along with varying from discrete and embedded morphologies; its transits recently vanished altogether \citep{Aungwerojwit2024}. ZTF\,J1944+4557 likewise shows variable transit activity, with transits fully ceasing and returning roughly every 2 years \citep{Guidry2025}. When transits are observable at ZTF\,J1944+4557, only the ingresses and egresses align at the 4.9704\,hr orbital period, with the in-transit morphology always changing from one cycle to the next. The periodic patterns of the relentless transits at ZTF\,J0328$-$1219 and WD\,1054$-$226 change on a timescale of weeks-to-months \citep{Vanderbosch2021,Farihi2022}; our observations show variations at \obj can be as rapid as one orbital cycle. All of this likely suggests some combination of dynamical processes at play, namely collisions, and possibly even differential fragmentation and streaming or interactions from an external, shepherding body. 

\section{Conclusions and Outlook}\label{sec:conclusion}

We collected over 176\,hr of follow-up, high-speed, time-series photometry in 2024~November and 2025~autumn that demonstrate \obj is still being continuously transited by planetary debris. Transits recurred exclusively at $P_{\rm A} = 9.937$\,hr during 2024~November, matching the period measured by TESS during Sectors~4 and 31. The previously transient period $P_{\rm B} = 11.17\,$hr began to dominate the line of sight in 2025~October; by 2025~November $P_{\rm A}$ was again predominant. Our conditional entropy periodogram detects a weak local minimum at 7.69\,hr during the final six days of our 2025~autumn observing season, which we connect to a potentially repeating transit at a new period of $P_{\rm C} =7.64$\,hr captured by our RVO light curves from October~29-November~1. Altogether, our light curves from both observing seasons illustrate the stability of $P_{\rm A}$ and also rapidly variable periodicity and transit morphology in this system, which we attribute to collisions.

We obtained an NUV spectrum with STIS over five consecutive HST orbits in 2025~October. We paired this spectrum with a new optical spectrum also obtained in 2025~October using SOAR/Goodman to update the model spectrum of \obj. Our NUV spectrum illuminates several metal lines from accreted material, including carbon, whose relative abundance ratios support the accretion of a chondritic object.

We generated an NUV light curve from our time-tagged HST/STIS spectrum to compare against near-simultaneous optical, ground-based photometry, measuring transits constant in color: $E(\mathrm{\textsc{nuv}} - \mathrm{\textsc{optical}}) / A_\mathrm{\textsc{optical}} = -0.10 \pm 0.12$. This result matches our measurement of likely colorless transits across our simultaneous $ugr$ photometry with ULTRACAM in 2024~November. Only during the fifth and final HST orbit do we measure a significant reddening term ($E(\mathrm{\textsc{nuv}} - \mathrm{\textsc{optical}}) / A_\mathrm{\textsc{optical}} = 0.59 \pm 0.08$). However, the transits during this fifth orbit are the most normalization and noise dominated among the dataset.

Our work like others before it \citep[e.g.,][]{Aungwerojwit2024,Korth2026_WD1054} encapsulates the importance of long-term, dedicated monitoring of transiting debris at white dwarfs. We also affirm the power of panchromatic astronomy. Measuring critical dust properties within white dwarf debris disks requires broad-band sampling of the SED \citep[e.g.,][]{Xu2019a,Farihi2025,Rogers2025a}. These results will be complemented by a forthcoming tandem study analyzing near- and mid-infrared observations of \obj collected in 2026~February by JWST DDT Program \#12507.

\section{Acknowledgments}

We are grateful for the constructive feedback offered by the anonymous reviewer that improved the communication of our analysis and findings. Discussions during the weekly Boston University White Dwarf Group meetings were critical to synthesizing these results and brainstorming how to most effectively communicate them. JAG acknowledges Ciarra Coston and Reuben Dsouza for their roles in these meetings. Colleagues at the 24th European Workshop on White Dwarfs shared similarly beneficial insight. The authors are indebted to all the observatory staff and telescope operators who made the collection of these light curves possible. We particularly thank Colt Pauley (PTO), Beck Maier, Ishara Nisley, Olivia Weiss (LDT), and C{\'e}sar Brice{\~n}o (SOAR).

This material is based upon work supported by the National Aeronautics and Space Administration under Grant No. 80NSSC23K1068 issued through the Science Mission Directorate.

This research is based on observations made with the NASA/ESA Hubble Space Telescope obtained from the Space Telescope Science Institute, which is operated by the Association of Universities for Research in Astronomy, Inc., under NASA contract NAS 5–26555. These observations are associated with program GO \#17789. The data described here may be obtained from the MAST archive at
\dataset[doi:10.17909/rye3-py98]{https://dx.doi.org/10.17909/rye3-py98}.

This project has received funding from the European Research Council (ERC) under the Europe an Union's Horizon 2020 research and innovation programme (Grant agreement No. 101020057).

These results made use of the Lowell Discovery Telescope (LDT) at Lowell Observatory.  Lowell is a private, non-profit institution dedicated to astrophysical research and public appreciation of astronomy and operates the LDT in partnership with Boston University, the University of Maryland, the University of Toledo, Northern Arizona University and Yale University.

This research is based on observations made with the Perkins Telescope Observatory, an astronomical observatory located at Anderson Mesa Station, Arizona, and is owned and operated by Boston University.

Based upon observations carried out at the Observatorio Astron{\'o}mico Nacional on the Sierra San Pedro M{\'a}rtir (OAN-SPM), Baja California, M{\'e}xico. We thank the daytime and night support staff at the OAN-SPM for facilitating and helping obtain our observations.

This work makes use of data from the Warwick 1m telescope operated on the island of La Palma by the University of Warwick at the Observatorio del Roque de los Muchachos of the Instituto de Astrof{\'i}sica de Canarias.

The contributions of C.-H.~B., X.-J.~J., H.-J.~W., J.-Q.~W., and T.-H.~Z. were supported by the Urumqi Nanshan Astronomy and Deep Space Exploration Observation and Research Station of Xinjiang (XJYWZ2303) and the Strategic Priority Research Program of the Chinese Academy of Sciences, Grant No. XDB0550100.

The SOAR/Goodman observations were conducted as part of program 2025B-136460, obtained through the Astronomical Event Observatory Network (AEON), a joint endeavor of the Las Cumbres Observatory and of NSF NOIRLab, which is managed by the Association of Universities for Research in Astronomy (AURA) under a cooperative agreement with the U.S. National Science Foundation.

This work has made use of data from the European Space Agency (ESA) mission
{\it Gaia} (\url{https://www.cosmos.esa.int/gaia}), processed by the {\it Gaia}
Data Processing and Analysis Consortium (DPAC,
\url{https://www.cosmos.esa.int/web/gaia/dpac/consortium}). Funding for the DPAC has been provided by national institutions, in particular the institutions
participating in the {\it Gaia} Multilateral Agreement.

J.A.G. attests generative artificial intelligence and large language models were not employed for any component of this analysis nor for the preparation of this manuscript.

\vspace{5mm}
\facilities{Gaia, HST (STIS), LDT (LMI), McD:Struve (ProEM), NTT (ULTRACAM), OANSPM:HJT, Perkins (PRISM), SO:Kuiper, SOAR (Goodman)}

\software{Astropy \citep{astropy:2013, astropy:2018, astropy:2022}, astroquery \citep{2019AJ....157...98G}, ccdproc \citep{matt_craig_2017_1069648}, extinction \citep{exctinction}, GCE \citep{Katz2021}, hipercam \citep{Dhillon2021}, IRAF \citep{1986SPIE..627..733T}, lmfit \citep{2016ascl.soft06014N}, matplotlib \citep{Hunter:2007}, nifty-ls \citep{2024ascl.soft12008G}, numpy \citep{harris2020array}, pandas \citep{mckinney-proc-scipy-2010,the_pandas_dev_team_2024_13819579}, phot2lc \citep{phot2lc}, photutils \citep{larry_bradley_2024_13989456}, pypeit \citep{Prochaska2020}, PyRAF \citep{PyRAF2012}, Pyriod \citep{2022ascl.soft07007B}, scipy \citep{2020SciPy-NMeth}, stistools \citep{stistools}, stenv (\url{https://stenv.readthedocs.io/}, wdwarfdate \citep{Kiman2022})
}

\appendix

\section{Observing Log of High-Speed, Time-Series, Follow-up Photometry}\label{sec:appendix_obs_log}

Table~\ref{tab:phot_obs} contains our observing log for the light curves described in Section~\ref{sec:observations} and shown in Figures~\ref{fig:UCAM_Nov2024_pfold_LCs} and \ref{fig:Oct2025_pfold_LCs}. Not included here is our attempt to obtain an $r$-band light curve the night of 2025~October~25, which would have been concurrent with our HST observations, using the Mookodi instrument on the 1-m Lesedi telescope at the South African Astronomical Observatory. Inclement weather prevented us from collecting photometry.

\begin{deluxetable*}{lccccccc}[!hb]
    \tablenum{A1}
    \tablecaption{Journal of Time-Series Photometry Observations of \obj.}
    \label{tab:phot_obs}
    \tabletypesize{\footnotesize}
    \tablewidth{0pt}
    \tablehead{
    \colhead{Date Obs} & \colhead{Telescope} & \colhead{Filter} & \colhead{$N_\mathrm{exp}$} & \colhead{$t_\mathrm{exp}$} & \colhead{Duration} & \colhead{FWHM} & \colhead{P2P}\\
    \colhead{(UTC)} & \colhead{} & \colhead{} & \colhead{} & \colhead{(s)} & \colhead{(hr)} & \colhead{(\arcsec)} & \colhead{(\%)}
    }
    \startdata
    2024-11-02T02:57:33.104 & ULTRACAM & {\em g} & 6252 & 3 & 6.02 & - & 1.7\\
    2024-11-02T02:57:33.104 & ULTRACAM & {\em r} & 6252 & 3 & 6.02 & - & 2.1\\
    2024-11-02T02:57:36.651 & ULTRACAM & {\em u} & 2084 & 10 & 6.02 & - & 4.8\\
    2024-11-03T01:45:02.239 & ULTRACAM & {\em g} & 7531 & 3 & 7.25 & - & 1.7\\
    2024-11-03T01:45:02.239 & ULTRACAM & {\em r} & 7531 & 3 & 7.25 & - & 2.1\\
    2024-11-03T01:45:05.786 & ULTRACAM & {\em u} & 2511 & 10 & 7.25 & - & 3.8\\
    2024-11-04T01:56:51.445 & ULTRACAM & {\em g} & 7320 & 3 & 7.05 & - & 2.0\\
    2024-11-04T01:56:51.445 & ULTRACAM & {\em r} & 7320 & 3 & 7.05 & - & 2.4\\
    2024-11-04T01:56:54.992 & ULTRACAM & {\em u} & 2440 & 10 & 7.05 & - & 3.9\\
    2024-11-05T02:54:50.387 & ULTRACAM & {\em g} & 6297 & 3 & 6.06 & - & 1.5\\
    2024-11-05T02:54:50.387 & ULTRACAM & {\em r} & 6297 & 3 & 6.06 & - & 1.8\\
    2024-11-05T02:54:53.935 & ULTRACAM & {\em u} & 2099 & 10 & 6.06 & - & 3.7\\
    2025-10-18T08:51:38.00 & OAN-SPM & {\em BG40} & 171 & 60 & 3.82 & - & 1.5\\
    2025-10-18T18:17:26.83 & PAT & {\em g} & 117 & 45 & 1.82 & - & 7.2\\
    2025-10-18T23:57:16.36 & Warw & {\em BG40} & 1722 & 10 & 5.88 & - & 6.8\\
    2025-10-19T06:39:56.01 & McD & {\em BG40} & 1297 & 10 & 3.61 & - & 4.4\\
    2025-10-20T08:42:50.00 & OAN-SPM & {\em BG40} & 330 & 30 & 4.06 & - & 1.3\\
    2025-10-21T00:00:25.00 & Stew & {\em V} & 146 & 50 & 2.47 & - & 1.3\\
    2025-10-21T02:50:31.11 & Warw & {\em BG40} & 115 & 60 & 1.96 & - & 1.5\\
    2025-10-21T06:03:51.01 & PTO & {\em BG40} & 900 & 20 & 6.25 & 2.50 & 1.7\\
    2025-10-21T08:35:42.00 & OAN-SPM & {\em BG40} & 161 & 30 & 1.85 & - & 1.5\\
    2025-10-22T05:37:08.22 & PTO & {\em BG40} & 186 & 20 & 1.31 & 2.41 & 2.3\\
    2025-10-22T09:03:10.01 & PTO & {\em BG40} & 455 & 20 & 3.29 & 3.59 & 1.6\\
    2025-10-23T01:37:31.19 & Warw & {\em BG40} & 648 & 20 & 3.94 & - & 2.1\\
    2025-10-23T05:27:56.02 & PTO & {\em BG40} & 525 & 20 & 3.65 & 3.61 & 1.5\\
    2025-10-23T09:45:20.19 & PTO & {\em BG40} & 369 & 20 & 2.56 & 3.04 & 1.6\\
    2025-10-24T05:27:11.07 & PTO & {\em BG40} & 983 & 20 & 6.83 & 2.87 & 1.5\\
    2025-10-25T01:28:30.01 & Warw & {\em BG40} & 651 & 20 & 3.95 & - & 2.3\\
    2025-10-25T09:16:11.41 & LDT & {\em r} & 550 & 15 & 2.98 & 0.99 & 1.1\\
    2025-10-25T19:19:30.543 & HST & NUV & 192 & 60 & 6.73 & - & 4.7\\
    2025-10-25T20:04:41.808 & LAST & Open & 540 & 30 & 6.30 & - & 6.8\\
    2025-10-26T04:59:15.8 & HAO & Open & 96 & 121 & 3.45 & - & 3.7\\
    2025-10-26T05:18:24.5 & RVO & Open & 350 & 67 & 6.76 & - & 2.3\\
    2025-10-26T05:21:13.00 & LDT & {\em r} & 183 & 15 & 1.10 & 1.36 & 1.1\\
    2025-10-26T06:29:04.57 & LDT & {\em r} & 332 & 15 & 1.85 & 1.36 & 0.9\\
    2025-10-26T08:34:39.45 & LDT & {\em r} & 24 & 15 & 0.13 & 1.06 & 0.7\\
    2025-10-26T09:07:38.47 & LDT & {\em r} & 565 & 15 & 3.18 & 1.20 & 1.6\\
    2025-10-27T00:56:47.70 & Warw & {\em BG40} & 202 & 60 & 3.63 & - & 2.1\\
    2025-10-27T04:56:45.3 & HAO & Open & 80 & 111 & 2.84 & - & 4.1\\
    2025-10-27T05:25:20.8 & RVO & Open & 343 & 60 & 6.59 & - & 2.1\\
    2025-10-28T04:32:07.5 & HAO & Open & 72 & 111 & 2.57 & - & 4.6\\
    2025-10-28T05:35:46.9 & RVO & Open & 334 & 60 & 6.36 & - & 2.1\\
    2025-10-29T04:39:26.9 & HAO & Open & 55 & 111 & 2.01 & - & 3.3\\
    2025-10-29T05:22:13.4 & RVO & Open & 343 & 60 & 6.52 & - & 1.9\\
    2025-10-30T05:02:10.6 & HAO & Open & 77 & 111 & 2.59 & - & 5.3\\
    2025-10-30T06:22:14.1 & RVO & Open & 194 & 60 & 3.70 & - & 1.7\\
    2025-10-31T05:07:39.5 & RVO & Open & 292 & 60 & 5.59 & - & 1.9\\
    2025-10-31T05:45:58.7 & HAO & Open & 30 & 111 & 1.33 & - & 5.0\\
    2025-11-01T05:05:23.1 & RVO & Open & 292 & 60 & 5.56 & - & 1.9\\
    2025-11-02T05:21:06.5 & RVO & Open & 286 & 60 & 5.72 & - & 2.1\\
    2025-11-03T05:22:55.2 & RVO & Open & 533 & 20 & 5.65 & - & 2.7\\
    \enddata
    \tablecomments{Telescope key names are the same defined in Section~\ref{sec:photfollowup}. We approximate the average seeing using FWHM measurements taken from the {\tt hipercam} pipeline outputs. P2P is the average point-to-point scatter in \% used to adjust the normalization of our light curves (Section~\ref{sec:lc_norm})}
\end{deluxetable*}

\clearpage
\onecolumngrid
\section{Supplemental 2025~October Light Curve Representations}\label{sec:appendix_lc_folds}

We supplement Figure~\ref{fig:Oct2025_pfold_LCs} by folding all our 2025~autumn light curves on $P_{\rm A}$ in Figure~\ref{fig:fold_A_Period} and the RVO and HAO light curves from October~29--November~3 on $P_{\rm C}$ in Figure~\ref{fig:fold_C_Period}. We plot our light curves unfolded in Figure~\ref{fig:lc_vs_time} to illustrate transit activity over the course of our campaign.

\renewcommand{\thefigure}{B.\arabic{figure}}
\setcounter{figure}{0}
\onecolumngrid
\begin{figure*}[!b]
    \epsscale{1.02}
\plotone{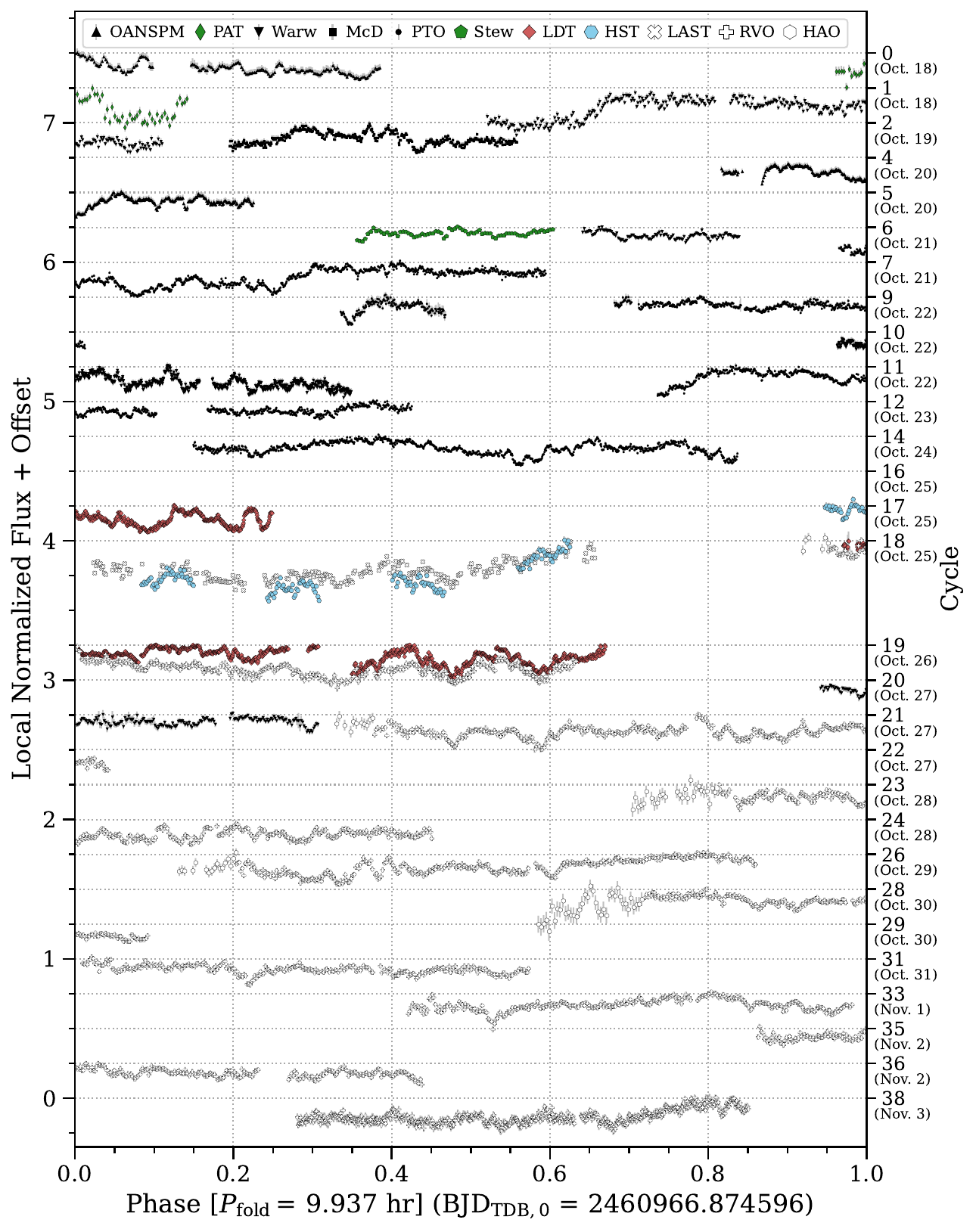}
    \caption{Light curves folded on $P_{\rm A} = 9.937\,$hr. It appears possible that Cycles 14 and 16 and Cycles 19 and 22 show evidence for being influenced by transits from this orbit. All light curves are plotted in the same fashion as that of Figure~\ref{fig:Oct2025_pfold_LCs}.
    }\label{fig:fold_A_Period}
\end{figure*}

\onecolumngrid
\begin{figure*}[!h]
    \epsscale{1.16}
     \plotone{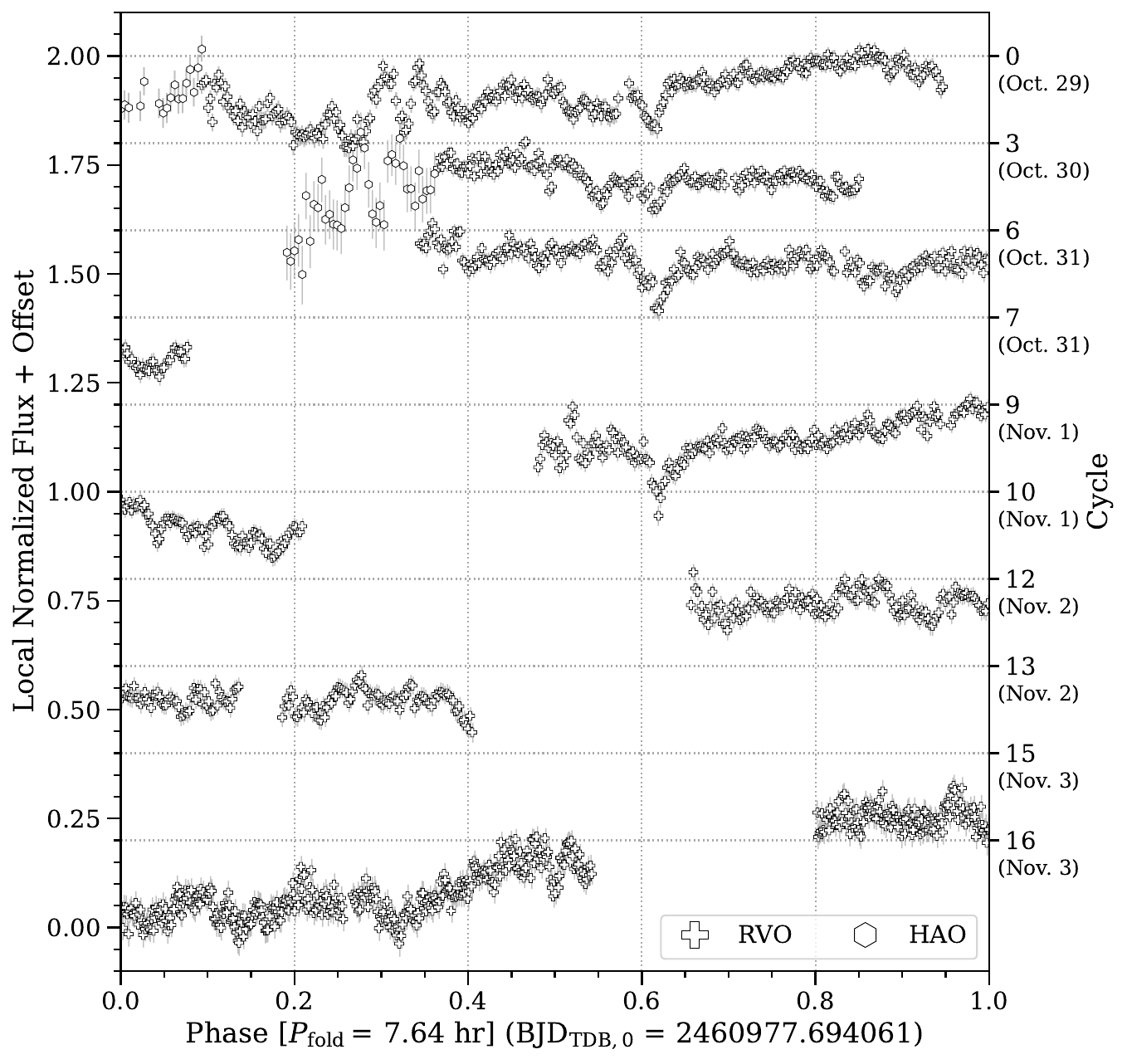}
    \caption{RVO and HAO light curves from 2025~October~29-November~3, the same grouping of light curves used for the third window in the bottom panel of Figure~\ref{fig:periodograms},  folded on $P_{\rm C} = 7.64\,$hr. A transit recurs through Cycles 0-9 at phase$\,\approx0.6$, despite conditional entropy indicating 7.69\,hr is the best period. Our RVO light curves during the nights of November~2 and 3 lacked the phase coverage to resolve the transit further.
    }\label{fig:fold_C_Period}
\end{figure*}

\onecolumngrid
\begin{figure*}[!h]
    \epsscale{1.12}
    \plotone{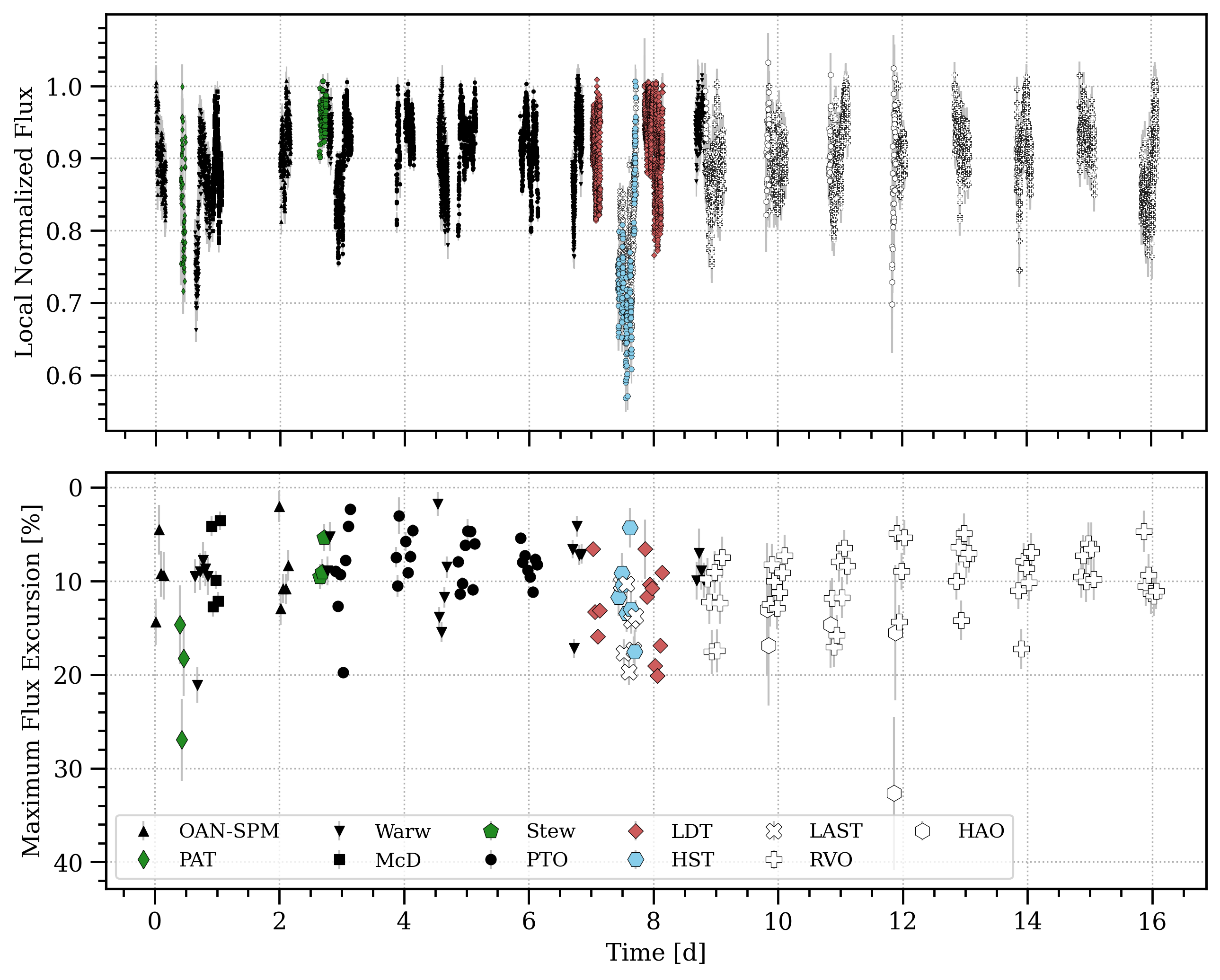}
    \caption{Our entire collection of 2025~October-November follow-up light curves (top) and the maximum total flux excursions computed over 1-hr bins of each light curve (bottom). Our HST observations notably occurred during the point of greatest activity we sampled. All light curves are plotted in the same fashion as that of Figure~\ref{fig:Oct2025_pfold_LCs}.
    }\label{fig:lc_vs_time}
\end{figure*}

\clearpage
\onecolumngrid
\section{2024~November ULTRACAM Light Curve Color Term Analysis}\label{sec:appendix_ucam_color_analysis}

We show our multi-color $ugr$ ULTRACAM light curves collected in 2024~November in Figure~\ref{fig:ucam_color_dependence}, along with the color term distributions we analyze in Section~\ref{sec:ucam_colors}. None of the color terms are definitively statistically significant when compared against the null hypotheses of horizontal, zero-slope distributions.

\renewcommand{\thefigure}{C.\arabic{figure}}
\setcounter{figure}{0}

\begin{figure*}[!h]
    \epsscale{1.12}
    \plotone{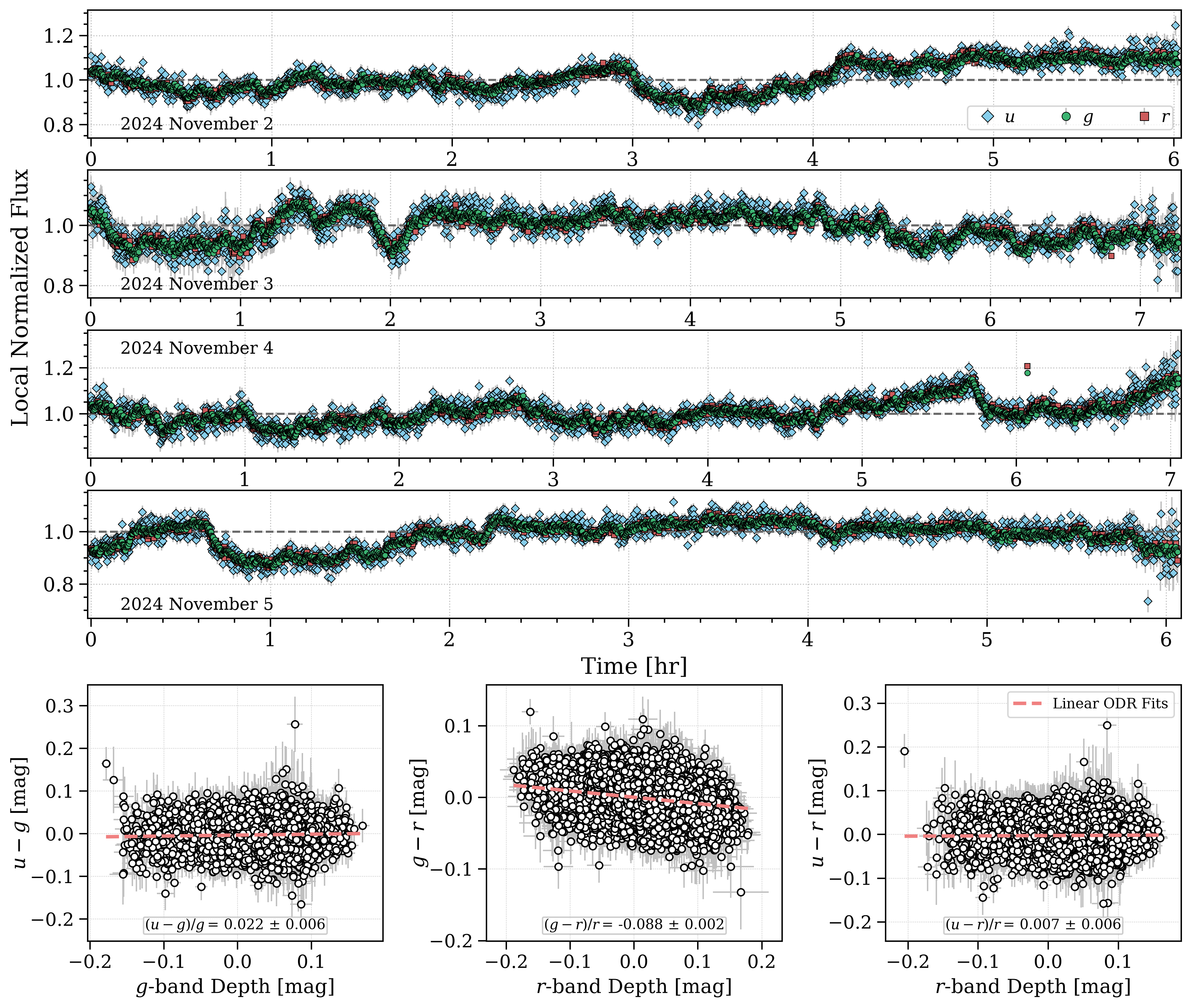}
    \caption{ULTRACAM $ugr$ light curves from 2024~November (top panels) and associated color term distributions. Despite the apparently significant color terms from $u$ vs. $g$ and $g$ vs. $r$, the residual variances of the linear ODR fits (salmon-colored lines) are indistinguishable from the null hypothesis of a horizontal line (i.e., colorless transits). See  Section~\ref{sec:ucam_colors}.
    }\label{fig:ucam_color_dependence}
\end{figure*}

\bibliography{references}{}
\bibliographystyle{aasjournal}

\end{CJK}
\end{document}